\pdfoutput=1  
\documentclass[preprint,authoryear,12pt]{elsarticle}

\usepackage{amsmath,amssymb}
\usepackage{graphicx}
\graphicspath{{figs/}{./}}
\usepackage{booktabs}
\usepackage{url}


\journal{Icarus}

\begin{document}

\begin{frontmatter}

\title{A calibrated dust-trail model of the Leonid meteoroid stream and
forecasts of the 2031--2035 encounters}

\author[aero,nusx]{Shinsuke Abe\corref{cor1}}
\ead{abe.shinsuke@nihon-u.ac.jp}
\cortext[cor1]{Corresponding author}
\affiliation[aero]{organization={Department of Aerospace Engineering,
  Nihon University}, addressline={7-24-1 Narashinodai},
  city={Funabashi}, postcode={274-8501}, state={Chiba}, country={Japan}}
\affiliation[nusx]{organization={Space Science Research Unit (NU-SX),
  Research Institute of Science and Technology, Nihon University},
  addressline={1-8-14 Kanda Surugadai}, city={Chiyoda-ku},
  postcode={101-8308}, state={Tokyo}, country={Japan}}

\begin{abstract}
We combine N-body dust-trail generation with a fast nodal-encounter
forecast and apply the model to the Leonid returns of 2031--2035.  The
dominant error is neither the ejection physics nor the encounter
machinery but the \emph{parent trajectory at the moment of ejection}: a
back-integration of 55P/Tempel--Tuttle places the pre-1700 ejection
sites more than 1~au from the fitted apparition solutions.  We therefore
anchor the ejection sites, not the dust, to the JPL Horizons ephemeris.
Anchoring removes an age-proportional bias in the modelled nodes and,
for the first time in this model, reproduces the classical attribution
of the 2001 and 2002 storms to the 1767 trail, with peak times accurate
to 1 and 40~min.  A second systematic is not a property of the parent: a
forecast read from a stored dataset holds each node where the snapshot
left it, while the real node moves a median $1.2\times10^{-2}$~au in
0.12 revolutions.  Two frozen snapshots of the same trail disagree by as
much, so each year uses its own epoch-matched dataset.  A completeness
audit --- every encounter the model predicts, not only the nominated
ones --- over-predicts the 1932 trail eighteenfold.  We predict two
comparable maxima on 2034 November 18, zenithal hourly rate
$\approx1.4\times10^{3}$ at 03:30~UT from the 1932 trail and
$\approx1.2\times10^{3}$ at 23:45~UT from the 1733 trail, with
$\approx1.4\times10^{3}$ in 2033 and $\approx650$ in 2035.  Which
dominates is a direct observational test of the trail-density term.
\end{abstract}


\end{frontmatter}

\section{Introduction}
\label{sec:intro}

The recognition that meteor storms are produced by narrow, young dust
trails rather than by the broad annual stream transformed shower
prediction from a statistical to a deterministic problem.
\citet{kondrateva1985} first computed the Leonid trail geometry by varying
only the ejection semi-major axis at perihelion;
\citet{mcnaught1999} and \citet{asher1999} developed this into the
trail-encounter formalism --- the radial miss distance $r_{\rm E}-r_{\rm
D}$, the ejection-velocity/radiation-pressure offset $\Delta a_0$, and the
mean-anomaly dilution factor $f_{\rm M}$ --- that predicted the 1999--2002
Leonid storms with timing errors of minutes.  \citet{lyytinen2000}
obtained comparable success with a radiation-pressure-dominated model and
a generalized-Lorentzian trail cross-section, and full numerical stream
models weighting simulated particles by cometary dust production
\citep{vaubaillon2005,egal2019,egal2020} now constitute the operational
state of the art; \citet{vaubaillon2005b} applied that scheme to the
Leonids themselves, predicting activity from very old (1466 and 1533)
trails in 2009 that \citet{koten2011} then tested against a
double-station video campaign, with photographic fireball data from the
same return reported by \citet{kokhirova2011}.  Reviews of two decades of predictions
\citep{egal2020review} conclude that peak \emph{times} are routinely
predicted to $\pm 0.5$~h, whereas peak \emph{rates} remain uncertain by a
factor of 2--3 in the best-calibrated cases and by 1--2 orders of
magnitude for poorly constrained parents.

The Leonids are the natural benchmark for any such model: five documented
storms (1833, 1866, 1966, 1999, 2001--2002) with well-measured rates and
times \citep{jenniskens2006}, a well-observed parent (55P/Tempel--Tuttle, $P \simeq 33.3$~yr)
with determined non-gravitational parameters, and a resonant filament
structure (the 5:14 Jovian commensurability;
\citealt{asher1999res}) that tests any model's treatment of trail aging.
With the next perihelion of 55P due on 2031 May 20, the 2030s encounters
are the next opportunity for storm-level Leonid activity, previously
examined by \citet{mcnaught1999} and \citet{maslov2007}.

This paper documents a dust-trail model developed for the interactive
simulator \emph{Meteorium}, and reports a systematic-error analysis that
led us to change how the model treats the parent body.  Epoch-matched
hindcasts of the 1999--2002 Leonid storms initially showed an outward
bias of the modelled nodal distances that grew in proportion to trail
age, together with a misattribution of the 2001/2002 storms to the 1833
trail rather than to the 1767/1866 trails of the classical analyses.  We
trace both to the accumulated error of the parent back-integration
(Sect.~\ref{sec:anchor}) and remove them by anchoring the ejection sites
to the JPL Horizons ephemeris.  Section~\ref{sec:stream} describes the
stream-generation model and the anchoring procedure,
Section~\ref{sec:forecast} the encounter forecast,
Section~\ref{sec:calib} the recalibration and validation, and
Section~\ref{sec:leonids} the 2030s Leonid predictions.

Five elements of what follows are, to our knowledge, new.
\emph{(i)} The ejection sites are anchored to the parent's
observation-fitted ephemeris rather than to a back-integration from a
single modern solution, and we show that this --- not the ejection
physics, the size distribution or the encounter kernel --- sets the
accuracy floor for models of this class, in a way that no internal
convergence test can detect.
\emph{(ii)} The hindcasts are scored in \emph{both} directions: every
maximum the model produces against every maximum the record holds.  That
audit exposes an eighteenfold rate error on one trail, and documented
maxima the model misses entirely, which a table restricted to the
nominated trails hides completely.
\emph{(iii)} The mean-anomaly factor is realized in absolute form,
normalized to the density a trail would have after one revolution of
unperturbed stretching, so that it can be compared directly against
published tabulations instead of only within one model.
\emph{(iv)} The population index is measured at each encountered section
from the local grain-size distribution rather than assumed constant, and
turns out to be corroborated trail by trail --- qualitatively by the
brightness annotations of an independent forecast, and quantitatively by
the only night on which the index was measured per trail: the two 2002
maxima differ by $0.33\pm0.11$ in the airborne television data
\citep{abe2003}, and the model puts them $0.31$ apart.
\emph{(v)} A second systematic of the same family as \emph{(i)} is
identified and removed.  Reading a forecast from a stream snapshot rather
than computing each encounter at its own epoch holds every trail's node
where the snapshot left it, while the real trail keeps being perturbed:
over a tenth of a revolution the nodes move $23\sigma_r$, and the omitted
motion is Jovian, not radiative.  Like the parent-trajectory error it is an error
of \emph{where the dust is put}, it is invisible to convergence tests,
and it was found only by comparing two computations that should have
agreed.  It reverses which trail dominates the 2034 encounter.

\section{Stream generation}
\label{sec:stream}

\subsection{Parent orbit and numerical integration}

The parent comet is initialized from its JPL Small-Body Database
osculating elements and integrated with the IAS15 integrator
\citep{rein2015} in \textsc{rebound} \citep{rein2012} together with the
eight planets.  Non-gravitational acceleration is applied to the parent
in the \citet{marsden1973} formulation,
\begin{equation}
\mathbf{a}_{\rm NG} = g(r)\,\bigl(A_1 \hat{\mathbf{r}} +
A_2 \hat{\mathbf{t}} + A_3 \hat{\mathbf{n}}\bigr), \qquad
g(r) = \alpha \left(\frac{r}{r_0}\right)^{-m}
\left[1+\left(\frac{r}{r_0}\right)^{n}\right]^{-k},
\end{equation}
with the standard water-ice constants ($r_0 = 2.808$~au, $m = 2.15$,
$n = 5.093$, $k = 4.6142$, $\alpha = 0.1113$) and the SBDB-fitted
$A_i$ (for 55P: $A_1 = 1.581\times10^{-9}$,
$A_2 = 9.186\times10^{-11}$~au\,d$^{-2}$).  The parent is first
integrated backward over the requested ejection era; perihelion passages
are located from the sampled heliocentric distance history, and each
passage defines one trail.

The 55P solution is fitted in eight parameters --- the cometary six
plus $A_1$ and $A_2$; $A_3$ and the perihelion-asymmetry delay $DT$ are
not fitted --- and Table~\ref{tab:cov} lists them with their formal
$1\sigma$ uncertainties.  The geometric elements are extremely well
determined, to parts in $10^{5}$ or better, and $t_p$ to $3.4$~s over a
132.5-yr arc; the non-gravitational terms are not.  $A_1$, which acts
radially and largely averages out over an orbit, carries a $22\%$
formal error, whereas $A_2$ --- transverse, and therefore accumulating
directly in the orbital period where a 132-yr arc of astrometry
constrains it tightly --- is fixed to $0.2\%$.  This asymmetry matters
for the present work because it is the period, not the instantaneous
acceleration, that governs where a trail ejected $n$ revolutions ago
lies today.  The parameters are also strongly correlated ($e$--$q$ at
$-0.999$, $A_1$--$A_2$ at $+0.941$, $e$--$A_2$ at $+0.844$,
$\Omega$--$i$ at $-0.804$), so they cannot be perturbed independently;
for this reason the full covariance matrix rather than its diagonal is
retrieved from the SBDB and archived with the parent element set (see
Data availability), and the ensemble of
Sect.~\ref{sec:systematics} draws from it.

\begin{table}[t]
\centering
\small
\caption{The eight fitted parameters of the 55P/Tempel--Tuttle orbit
solution and their formal $1\sigma$ uncertainties, from the JPL SBDB
covariance (solution J985/69, epoch JD 2451040.5, 392 optical
observations over a 132.5-yr arc, fit RMS $0\rlap{.}''73$).  The set is the
standard cometary six --- $e$, $q$, $t_p$, $\Omega$, $\omega$, $i$ ---
plus the two Marsden non-gravitational parameters; $A_3$ and the delay
$DT$ are not fitted for this comet.  The geometric elements are
determined to parts in $10^{5}$ or better and $t_p$ to 3.4~s, but $A_1$
carries a 22\% formal error while $A_2$, which acts directly on the
orbital period and is therefore visible across the whole arc, is
determined to 0.2\%.  Strong correlations ($e$--$q$ at $-0.999$,
$A_1$--$A_2$ at $+0.941$, $e$--$A_2$ at $+0.844$, $\Omega$--$i$ at
$-0.804$) mean the parameters cannot be varied independently, which is
why the full matrix is carried rather than the diagonal.  The matrix is
reproduced in full below rather than summarised, so that the ensemble
of Sect.~\ref{sec:systematics} can be redrawn from the paper alone.}
\label{tab:cov}
\begin{tabular}{llll r}
\toprule
Parameter & Value & $1\sigma$ & Unit & Relative \\
\midrule
e & 0.905553 & 6.803e-08 &  &  \\
q & 0.976428 & 7.348e-07 & au &  \\
tp & 2.45087e+06 & 3.901e-05 & d &  \\
node & 235.2710 & 1.001e-04 & deg &  \\
peri & 172.5003 & 1.460e-04 & deg &  \\
i & 162.4866 & 1.932e-05 & deg &  \\
A1 & 1.58098e-09 & 3.486e-10 & au\,d$^{-2}$ & 22.1\% \\
A2 & 9.18642e-11 & 1.974e-13 & au\,d$^{-2}$ & 0.2\% \\
\bottomrule
\end{tabular}
\end{table}

\subsection{Ephemeris anchoring of the ejection sites}
\label{sec:anchor}

\begin{figure}[t]
\centering
\includegraphics[width=0.74\linewidth]{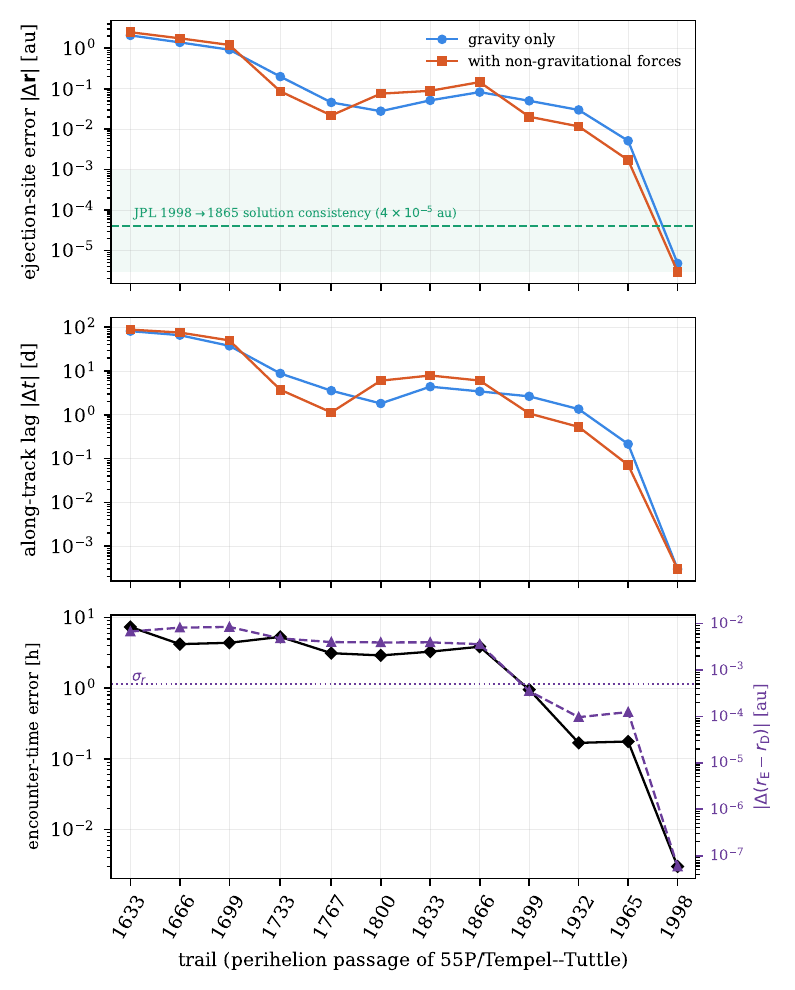}
\caption{Error of the internal parent back-integration at the ejection
site of each Leonid trail, measured against the JPL Horizons ephemeris
of 55P/Tempel--Tuttle.  \emph{Top}: heliocentric position error at
perihelion passage, for the gravity-only integration (blue circles) and
with Marsden non-gravitational forces (red squares); the dashed line and
shaded band mark the level to which JPL's own integration of the 1998
orbit solution reproduces the independently fitted 1865 apparition
solution.  \emph{Middle}: the same error expressed as an along-track
lag.  Both integrations start from the current (1998) orbit solution, so
the error vanishes at the right-hand edge by construction and grows
monotonically into the past; the trails responsible for the historical
storms (1767, 1866, 1899, 1932) carry ejection-site errors of
$0.02$--$0.15$~au, and pre-1700 trails exceed 1~au.  Non-gravitational
forces do not systematically help: one modern $A_1, A_2$ pair cannot
represent three centuries of outgassing.  \emph{Bottom}: what that error costs the \emph{prediction}
--- the descending-node longitude of the orbit the dust inherits,
converted to an Earth-arrival time at $0.9856^\circ$~d$^{-1}$ (left,
black diamonds), and the node distance $r_{\rm E}-r_{\rm D}$ it inherits
(right, violet triangles; the dotted line is $\sigma_r$).  Both are first-order errors
seeded at ejection, before three centuries of dust propagation act on
them.}
\label{fig:anchor}
\end{figure}

A dust trail inherits the parent's state at the instant of ejection, so
an error in the historical parent trajectory displaces the entire trail
rigidly.  This error is not negligible.  Figure~\ref{fig:anchor}
compares our backward integration of 55P --- started from the
current (1998-apparition) orbit solution and run with the full planetary
force model, with and without the non-gravitational terms --- against the
JPL Horizons ephemeris, i.e.\ against JPL's own integration of the
current (1998-apparition) orbit solution with their force model.  The error is negligible for the
1998 and 1965 trails ($<5\times10^{-3}$~au), reaches $0.03$--$0.05$~au
at the 1932 and 1899 trails that produced the 1966 and 1999 storms,
$0.08$~au (gravity only) or $0.15$~au (with non-gravitational forces) at
the 1866 trail, and exceeds 1~au for every trail ejected before 1700.
Expressed as an along-track lag --- the time by which the modelled comet
arrives early or late at its true position --- the same sequence runs
from $10^{-3}$~d (1998) through $1$--$8$~d (1767--1932) to 40--90~d
(1633--1699).  Because the discrepancy is almost entirely along-track,
it appears in the trail cross-section as a shift of the nodal crossing
along Earth's orbit and --- through the parent's changing heliocentric
distance at the displaced ejection point --- as a systematic offset of
$r_{\rm node}$ that grows with the number of revolutions since ejection.
It is not an artefact of our integrator, and the JPL track is not
merely a different guess: Horizons also holds orbit solutions fitted
separately to each historical apparition of 55P (1366, 1699, 1865, 1965,
1998), and the back-integrated 1998 solution reproduces the
independently fitted 1865 solution at the 1866 perihelion to
$1\times10^{-5}$~au (1965: $2\times10^{-5}$~au).  What accumulates is the mismatch
between a single modern non-gravitational parameterization and the
apparition-to-apparition variability of a real comet's outgassing, a
limitation already noted for 21P by \citet{egal2019} and expressed as
an empirical $A_2$ jitter by \citet{lyytinen2001}.

We therefore separate the two jobs the parent integration performs.  The
\emph{ejection geometry} is taken from the Horizons ephemeris: for each
scheduled dust burst the parent's heliocentric ecliptic state is
interpolated (cubic Hermite, on a 1-d grid spanning $\pm900$~d around
each perihelion) from the tabulated ephemeris and used as the origin and
velocity reference for the released grains.  The \emph{dust propagation}
remains entirely within our own N-body integration, so radiation
pressure, Poynting--Robertson drag and planetary perturbations act on
the grains exactly as before.  Horizons integrates comets no earlier
than 1599 December 11; ejection eras before that date fall back to the
internal back-integration, and the fraction of anchored trails is
recorded with each dataset.  For the eighteen-return Leonid datasets used
here (1433--1998) every trail is anchored: the twelve returns from 1633
onward to the modern solution, the six before 1600 to the 1366-epoch
solution, which Horizons propagates below the modern one's floor.

\paragraph{What the site error costs the prediction}
An ejection-site error matters only through what the dust inherits from
it, and the two quantities that set a forecast are the node's
\emph{longitude}, which fixes the encounter time, and its \emph{distance}
from Earth's orbit, which fixes the rate.  Taking the parent state at
each perihelion from both sources and forming the osculating orbit of
each isolates the error the trail is seeded with, before any dust
propagation acts on it.  Converting the node-longitude difference at
Earth's mean motion ($0.9856^\circ$~d$^{-1}$) gives the bottom panel of
Fig.~\ref{fig:anchor}.

The result is not proportional to $|\Delta \mathbf{r}|$, and that is the
useful part.  A node longitude is an orientation, so it saturates: the
seeded timing error is $3$--$7$~h for every trail from 1633 to 1866,
whether the site is misplaced by $0.02$~au (1767) or by $2.5$~au (1633).
It falls below an hour only from 1899 onward --- $0.96$~h (1899),
$0.17$~h (1932), $0.18$~h (1965) --- and vanishes at 1998 by
construction.  A small site error in an unfavourable direction is
therefore as damaging to the timing as a large one.

The node \emph{distance} behaves the same way and is the more
consequential of the two: the inherited $|\Delta(r_{\rm E}-r_{\rm D})|$
is $4$--$8\times10^{-3}$~au for the 1633--1866 trails, that is $8$--$16$
times the encounter kernel $\sigma_r$, and drops below $\sigma_r$ only
for 1899 and later.  A trail seeded eight kernel widths from its true
node is not merely mistimed; it is placed on the wrong side of the
profile, which is why the unanchored model attributed the 2001 and 2002
storms to the 1833 trail rather than to 1767.  Two independent checks
support the scale: the per-apparition experiment below, whose ejection
sites differ from the adopted ones by $0.04$--$0.11$~au, degrades the
measured peak times to $2.5$--$3.4$~h, matching the $\simeq3$~h this
calculation predicts for that displacement; and the residual node offset
that survives anchoring, $1$--$2\times10^{-3}$~au
(Sect.~\ref{sec:systematics}), is a factor four smaller than the seeded
error it replaced.

\subsection{Ejection model}
\label{sec:ejection}
Every model in this class descends from \citet{whipple1951}, who first
showed that the sublimating ices of a cometary nucleus drag embedded
solid grains outward and release them with a terminal speed set by the
gas flux, the grain size and the nucleus radius --- the mechanism that
turns a comet into a meteoroid stream, and the reason a trail has a
finite width at all.  The forms used below are modern parameterizations
of that picture.
Grains are released while $r < 3$~au, at burst times sampled proportional
to a dust-production law $\propto r^{-n_{\rm prod}}$ along each perihelion
arc.  Each grain receives a physical radius $a$ drawn directly from the
differential size distribution $\mathrm{d}N \propto a^{-u}\,\mathrm{d}a$
over $[a_{\min}, 2\,\mathrm{cm}]$.  The lower truncation
\begin{equation}
a_{\min} = \max\!\left(a_{+6.5},\;
\frac{5.74\times10^{-4}}{\rho\,\beta_{\max}}\right)
\label{eq:amin}
\end{equation}
combines (i) the smallest grain that produces a $+6.5$-mag meteor at the
shower's atmospheric speed, obtained by inverting the
\citet{jacchia1967} mass--magnitude--velocity relation
$\log_{10} m\,[\mathrm{g}] = 6.06 - 0.62\,M_v - 3.89 \log_{10}
v\,[\mathrm{km\,s^{-1}}]$, and (ii) an upper bound
$\beta_{\max} = 3\times10^{-3}$ on the radiation-pressure parameter,
reflecting the empirical composition of storm-producing trail sections
($\Delta a_0 \simeq +0.2$~au $\Leftrightarrow \beta \simeq 10^{-3}$ for
the Leonids; \citealt{mcnaught1999,asher1999}).  Because sizes are
sampled from the distribution itself, and ejection times from the
production law, \emph{every simulated grain represents exactly one visual
meteoroid and carries unit statistical weight}; this pure-Monte-Carlo
design keeps the forecast robust at the moderate particle counts
($10^2$--$10^3$ per trail) required for interactive use, in contrast to
importance-weighting schemes that demand $10^6$--$10^8$ particles
\citep{vaubaillon2005b,egal2019}.  The two designs answer different
questions: a weighted ensemble carries the parent's absolute dust
production and can predict a flux without a storm calibration, whereas
ours carries only the geometry and buys its amplitude from one measured
storm per shower.
The radiation-pressure parameter follows physically from the size,
$\beta = 5.74\times10^{-4} Q_{\rm pr}/(\rho a)$ (SI;
\citealt{burns1979}), with $Q_{\rm pr}=1$ and per-shower bulk density
$\rho$ (400~kg\,m$^{-3}$ for the Leonids).  Terminal ejection speeds use
the \citet{crifo1997} parametric form as implemented by
\citet{vaubaillon2005} and \citet{egal2019},
\begin{equation}
v(a, \theta, r) = k_{\rm ej}\,
\frac{W}{1.2 + 0.72\sqrt{a/(a_\ast \cos\theta)}}, \qquad
W = \sqrt{\tfrac{\gamma+1}{\gamma-1}\,\gamma k_{\rm B} T / m_{\rm g}}
\simeq 930~\mathrm{m\,s^{-1}},
\label{eq:cr97}
\end{equation}
where $\theta$ is the solar zenith angle of the (cosine-weighted,
sunlit-hemisphere) ejection direction, and the gas--grain coupling radius
$a_\ast \propto Q_{\rm gas}/(\rho W R_{\rm c})$ is scaled per comet
($a_\ast = 0.5\,\mu$m at 1~au for 55P, $\propto r^{-2}$) so that
millimetre grains leave at the 20--30~m\,s$^{-1}$ speeds required by the
Leonid storm calibrations \citep{mcnaught1999,asher1999} and consistent
with the low-speed tail measured in situ at 67P
\citep{dellacorte2015,agarwal2016}.  A global multiplier $k_{\rm ej}$
(default 1) accommodates fragmentation-driven events, for which
2--5$\times$ nominal speeds are required \citep{egal2023}, and
low-activity parents.  The \citet{brown1998} law, closer to \citeauthor{whipple1951}'s
original scaling and with a parabolic speed spread, is retained as an
alternative
(\texttt{bj98}); the two agree for the Leonids at the calibrated
$a_\ast$.  For the Leonid configuration the sampled speed distribution
has a median of 28~m\,s$^{-1}$ (10th--90th percentile 17--35~m\,s$^{-1}$)
at 1~au.
Grains are integrated forward to the snapshot epoch with radiation
pressure and Poynting--Robertson drag via per-grain $\beta$
(\textsc{reboundx}; \citealt{tamayo2020}), and stored as heliocentric
osculating elements together with $a$, $\beta$, and the trail (ejection
return) label.
\section{Encounter forecast}
\label{sec:forecast}
\subsection{Nodal kernel}
For a target year the forecast propagates each grain's frozen elements
with two-body motion and evaluates, for the ecliptic node nearest 1~au:
the longitude at which Earth (from the JPL ephemeris) meets the node, the
radial miss distance $\Delta r = r_{\rm node}-r_{\rm E}$, and the timing
offset $\Delta t$ between the grain's own nodal passage and Earth's
arrival.  The activity profile is evaluated on a 15-min solar-longitude grid,
fine enough to separate the contributions of overlapping trails.
The along-trail density at the encounter is estimated with an
adaptive $k$-nearest-neighbour kernel ($k=4$, floor
$\sigma_t = 15$~d, edge floor $|\Delta t|/3$), so that sparse sections of
old trails retain a small non-zero density while dense young trails stay
sharp.  The radial profile uses the generalized-Lorentzian cross-section
of \citet{lyytinen2000},
\begin{equation}
K(\Delta r) = \left[1 + (\Delta r/\sigma_r)^2\right]^{-1.35},
\end{equation}
whose wings match the observed storm profiles better than a Gaussian
\citep{jenniskens2000}; $\sigma_r$ is a per-shower constant
($5\times10^{-4}$~au for the Leonids, consistent with the
(2.4--2.7)$\times10^{-4}$~au Gaussian $\sigma$ fitted by
\citealt{mcnaught1999}).

Three conventions exist for deciding which grains take part in an
encounter.  \citet{asher1999imc} does not impose a time window at all,
because none is needed: he parameterizes the trail by the ejection
offset $\Delta a_0$ rather than by the timing offset $\Delta T$, noting
that ``$\Delta T$ immediately after ejection is zero for all
particles'' whereas $f_a(\Delta a_0)$ connects directly to an ejection
model.  Fixing the ejection return and the encounter year then fixes
the elapsed time, hence the orbital period, hence $a_0$ --- so that,
in \citeauthor{asher1999}'s words, ``only meteoroids with the right
mean anomaly $M$ to impact Earth at whatever date of interest need be
considered'' \citep{asher1999}.  The selection happens before the
integration, and the cross-section figures that result (his Fig.~8,
which Sect.~\ref{sec:t1932} compares against) plot only grains that
cross in the year of interest.

Weighted-ensemble models integrate forward instead and cut afterwards:
\citet{gockel2000} keep grains within
$L_{\rm lim} = 10^{-3}$~au and $T_{\rm lim} = 7$~d of Earth, and
\citet{vaubaillon2005b} select near-Earth grains on a spatial criterion
equivalent to $\Delta T \simeq 17.5$~h, then impose a second, one-hour
criterion on the grains treated as impacting.  Those cuts are
affordable at $10^{6}$--$10^{8}$ particles.  At the $10^{2}$--$10^{3}$
per trail this model carries they would leave most sections empty, so
we do not cut on $\Delta t$ at all: every grain contributes through the
kernels above, weighted by its radial miss distance and by its
along-trail timing offset.  This model therefore sits between the two
conventions --- it shares \citeauthor{asher1999imc}'s single $f_{\rm M}$
but, integrating every trail forward rather than solving for the $a_0$
that reaches Earth, it inherits the weighted-ensemble models' need for
a cut without adopting one.  The price is that a grain far out of phase
retains a small weight, and Sect.~\ref{sec:leo2009} shows what that
cost when the along-trail bandwidth was allowed to grow without bound.
The cross-section figures therefore mark the
\citeauthor{gockel2000} criterion explicitly, so that a reader can see
which part of each plotted section would survive the stricter
convention.
\subsection{Amplitude}
\label{sec:amplitude}
The ZHR amplitude of trail $i$ is
\begin{equation}
\mathrm{ZHR}_i = Z_{\rm storm}\; O_i\; f_{{\rm M},i}\; f_{{\rm disp},i},
\label{eq:amp}
\end{equation}
where $Z_{\rm storm}$ is a single per-shower calibration constant (the
rate of a direct hit on a dense one-revolution trail), $O_i$ is a
self-normalizing radial overlap factor that measures how much of the
trail's density-weighted cross-section lies on the Earth line relative to
the best case of the same cross-section slid onto the line (with an extra
crest-distance suppression for broad clouds), and $f_{{\rm M},i}$ is an
\emph{absolute} realization of the \citet{mcnaught1999} mean-anomaly
factor: the along-trail linear density of the encountered section,
$\rho_{\rm enc}$ (from the same kernel estimate that shapes the
profile), normalized by the density the same trail would possess after
exactly one revolution of unperturbed linear stretching,
\begin{equation}
f_{{\rm M},i} = \frac{\rho_{{\rm enc},i}}
{N_i / (\sqrt{2\pi}\,\sigma_{P,i})},
\label{eq:fm}
\end{equation}
with $N_i$ the trail's grain count and $\sigma_{P,i}$ the robust spread
of its grains' orbital periods --- a quantity the simulation carries
intrinsically.  Equation~(\ref{eq:fm}) reproduces
$f_{\rm M} \simeq 1/n_{\rm rev}$ to first order, retains the full
positional dependence caused by planetary perturbations, and --- unlike a
dataset-relative reference --- is invariant to the dataset's trail
composition (verified: removing fourteen of eighteen trails leaves the
2034 encounter amplitude unchanged to five significant digits).  The
remaining age term
\begin{equation}
f_{{\rm disp},i} =
\begin{cases}
\exp[-(n_{\rm rev}-12)/3], & n_{\rm rev} > 12 \text{ and non-resonant},\\
1, & \text{otherwise},
\end{cases}
\end{equation}
carries only the trail-coherence limit of \citet{asher1999imc}.  Critically, the
cutoff is \emph{waived for resonant trails}: a trail whose period lies
within 1.5\% of a $p\!:\!q \le 15$ commensurability with Jupiter is
protected, preserving structures such as the Leonid 5:14 filament (the
1998 fireball display originated in the 1333 trail;
\citealt{asher1999res}) and the 1:6 Halleyid clusters
\citep{sato2007,egal2020}.  An IMO-style annual background,
$\mathrm{ZHR}_{\rm bg} = Z_{\rm cat}\,10^{-B|\lambda_\odot -
\lambda_{\max}|}$ with $B = 0.2$~deg$^{-1}$, is added for theory
datasets, and each grain is counted only if its Jacchia magnitude at the
shower speed is brighter than $+6.5$.
\subsection{Limiting magnitude and the population-index conversion}
\label{sec:maglim}

A ZHR is \emph{defined} as the rate a single observer would record with a
limiting magnitude of $+6.5$, but visual and video observers detect only
to about $+4$; a published ZHR is therefore the detected count
extrapolated to $+6.5$ with the population index $r$.  With $r \simeq 2$
that extrapolation is a factor $r^{2.3} \simeq 5$, so most of any
reported ZHR is extrapolation rather than detection, and the extrapolation
assumes a single $r$.

Our size sampler starts at the largest of the absolute floor, the
$+6.5$-magnitude grain and the radiation-pressure cut
$a = 5.74\times10^{-4}/(\rho\,\beta_{\rm max})$.  For the Leonids the
$\beta$ cut wins: the smallest simulated grain is $0.478$~mm, which by
\citet{jacchia1967} at $71$~km~s$^{-1}$ is a $+4.2$~magnitude meteor.
The model therefore counts \emph{detected} meteors, and its magnitude cut
never fires --- 100.0\% of grains clear $+6.5$ in every section of every
year examined here.  That the sampler floor lands within $0.2$~mag of the
practical detection limit is a coincidence of the adopted $\beta$ cap, but
a convenient one: it means the simulated population is the observed one.

The conversion to a ZHR is then explicit.  Radiation pressure sorts a
trail by grain size along its length, so the section Earth meets is not a
fair sample of the trail, and we measure the local size distribution at
every encounter --- a Pareto maximum-likelihood slope
$u = 1 + n/\sum\ln(a_i/a_{\min})$ for
$\mathrm{d}N \propto a^{-u}\,\mathrm{d}a$, giving the mass index
$s = (u+2)/3$ and $r = 10^{(s-1)/2.5}$ --- and convert with that section's
own $r^{6.5 - 4.2}$.  The measured slopes are strongly sorted and the
sorting correlates with the model's performance: the 1767 sections it
reproduces best are bright-enriched ($u = 1.9$--$2.2$, $r = 1.33$--$1.44$)
while the 1932 sections it over-predicts are faint-enriched
($u = 2.9$--$3.6$, $r = 1.78$--$2.15$), against $u \simeq 3.5$,
$r \simeq 2.15$ for every trail taken whole.  This is the mass sorting
\citet{lyytinen2000} invoke when they annotate their 2000 November 17 row
\emph{``mostly faint''}, and for one night it has been measured directly.
\citet{abe2003} determine the magnitude distribution index separately for
the two 2002 maxima from the airborne television data, and find
$1.70\pm0.10$ on the seven-revolution 1767 peak at 04:03~UT against
$2.03\pm0.05$ on the four-revolution 1866 peak at 10:49~UT --- the first
rich in bright meteors including fireballs, the second not.  Our sections
give $1.44$ and $1.75$.  Both are low by $\simeq0.27$, an offset the storm
calibration absorbs because it is common to the two, but the
\emph{difference} between the trails, $0.31$ against an observed
$0.33\pm0.11$, is reproduced to $0.02$.  The two determinations are not
taken over the same magnitude range --- theirs is restricted to
$m \le +3.5$, ours runs over everything the sampler carries --- so the
offset is not to be read as a discrepancy; the contrast is the test, and
it is a per-trail one on the quantity this model computes.  We clip $r$ to the observed Leonid range
$[1.3, 3.0]$, and the measured slope to a ceiling $u_{\max}$ whose default
is the shower's own ejected index --- $3.5$ for the Leonids and Perseids,
$2.9$ for the Draconids, giving $r \le 2.15$ and $r \le 1.79$
respectively.
The second bound matters more than it sounds.  The estimator is a Pareto
MLE above the dataset's smallest grain, so a section whose own grains sit
against that floor --- which is what a wing encounter selects, radiation
pressure having placed the small-grain edge of the trail on Earth's orbit
--- returns a slope fixed by the truncation rather than by the
population: the 1965 section of 2001 reads $u = 11.1$ with its median
grain 10\% above the floor.  The model has nothing below its own floor to
extrapolate from, so such a slope must not become an $r^{2.3}$
amplification.  The separation is clean: every section that reaches the
bound has its median grain within $1.10$--$1.44$ floors, and every
section reproducing an observed storm spans $1.60$--$2.59$.  Imposing it
leaves every encounter in Table~\ref{tab:leo9802} that the record
contains unchanged, and removes up to a factor two from those it does
not.

Scored fairly --- re-deriving $Z_{\rm storm}$ for each configuration, so
that the test measures shape and not scale --- making the conversion per
section improves six of the nine matched comparisons of
Table~\ref{tab:leo9802} and worsens three.  The split is not random.
Every comparison in which the model is within a factor two of the record
improves, the largest single gain being the 2000 1932 section
($\times0.35 \to \times1.00$); the three that worsen are the 1733 and
1866 sections of 2000 and the 1800 section of 2001, all of them already
under-predicted by factors of 11--40, where the change is in the third
decimal place.  The conversion also makes the calibration constant
physically interpretable: without it the same fit returns
$Z_{\rm storm} = 1.8\times10^{4}$, the value earlier versions of this
model carried, because the constant was absorbing the magnitude
extrapolation implicitly.  It is adopted throughout.

\subsection{Residual systematics}
\label{sec:systematics}
Two systematics survive the anchoring of Sect.~\ref{sec:anchor}.  The
first is not a limitation of the stream integration --- which carries the
planets, radiation pressure and Poynting--Robertson drag on every grain
from its ejection to the epoch at which the dataset is written --- but of
how a stored dataset is used afterwards.  To forecast a year the model
reads that dataset and advances each grain with two-body motion, so only
its mean anomaly changes: the semi-major axis, eccentricity, inclination
and the two angles stay at their stored values, and with them the node.
A forecast computed away from the dataset's epoch therefore does not
misplace the node; it \emph{leaves it where it was}, while the real grain
carries on being perturbed.  The error is the motion omitted, and its
size is the question.

We first met it as a hindcasting rule --- the 1999 storm recovered at
only $\times0.07$ from the 2026-epoch dataset --- and assumed, as the
staleness guard of Sect.~\ref{sec:conf} did, that \emph{forward}
extrapolation was safe within a fraction of a revolution.  It is not.
Running one integration per return and writing it out at several epochs,
so that the ejection realisation
and the entire dust history up to the first snapshot are shared exactly,
we can measure the omitted motion directly: between the 2030 and the
2034 snapshot the nodes of all eighteen trails move in the same
direction, over an interval of 0.12 revolutions, by a median
$1.15\times10^{-2}$~au grain by grain --- $23\sigma_r$ --- from
$0.6\times10^{-3}$~au for the 1998 trail, still essentially where the
comet left it, to $1.9\times10^{-2}$~au for 1866.  Propagated instead
with frozen elements, none of that motion happens; what the forecast
then reports as the section's node is in error by $6.4\times10^{-3}$~au,
$13\sigma_r$, the smaller figure because the reported node is a median
over the grains the encounter selects.

The omitted motion is planetary, not radiative.  Integrating the same
grains from the 2030 snapshot to 2034 with the planets alone recovers
87--100\% of it (1932: $5.9$ of $6.8\times10^{-3}$~au; 1767: $17.69$ of
$17.69$; 1733: $12.08$ of $12.12$; 1899: $7.4$ of $7.6$), while the Sun
with radiation pressure and Poynting--Robertson drag but no planets
contributes $\le0.2\times10^{-3}$~au, a fraction of $\sigma_r$.  Jovian
perturbation does not stop acting on a trail because a dataset has been
written to disk.  That the full force set reproduces the independently
written 2034 snapshot to $0.3\sigma_r$ also confirms that the two
snapshots of one integration are consistent, which is what makes the
comparison a measurement rather than a discrepancy between two runs.
Two frozen snapshots of the same trail disagree with each other by as
much: the 1633 node is $+1.6\times10^{-3}$~au extrapolated from 2026 and
$+9.6\times10^{-3}$~au from 2030, against $+4.5\times10^{-3}$~au when the
N-body is carried to the encounter.  On the steep flank of the radial
profile that is the difference between a storm and nothing, and it is
not a small-number effect: it survives at $2.5\times10^{4}$ grains in the
trail.

The hindcasts of Sect.~\ref{sec:calib} never incurred this, because each
was already computed from a dataset snapshotted in its own encounter
year; it is the 2030s forecasts that were exposed, having been read from
a single 2026-epoch dataset.  Every forecast in this paper is now
computed at its own epoch as well, and the staleness guard is tightened
from 0.30 revolutions to zero: the forecast refuses any year but the
dataset's own.  The systematic is a property of snapshot-and-propagate
designs, not of trail models in general --- the weighted-ensemble
forecasts integrate their $10^{6}$--$10^{7}$ particles to each encounter
directly \citep{vaubaillon2005b,egal2019} and never incur it.  Ours took
the snapshot route because the model serves an interactive simulator,
where a forecast has to be produced in seconds from a stored stream; the
finding is therefore a caution for any model that reuses one snapshot
across years, which is what that requirement makes tempting.  The
construction that makes the alternative affordable --- one N-body
integration per return, snapshotted at each forecast year and merged
into one dataset per year --- reproduces the 1999 storm at $\times1.03$
with the node at $-0.89\times10^{-3}$~au, against $\times0.89$ and
$-0.87\times10^{-3}$~au for the independently built dataset of
Table~\ref{tab:leo9802}, so the merge itself introduces nothing.
Second, a small residual node-longitude offset remains for the oldest
trails.  Before anchoring, epoch-matched hindcasts of the 2001 and 2002
encounters required an empirical correction of $-0.25^\circ$ for trails
older than three revolutions; with the ejection sites anchored, the same
hindcasts are best reproduced with $-0.11^\circ$, i.e.\ anchoring removes
somewhat more than half of the offset.  We retain the reduced correction
as a per-dataset parameter applied only to trails with
$n_{\rm rev} \ge 4$, in the same spirit as the empirical $r_{\rm D}$
offsets of \citet{mcnaught1999}.  What remains is plausibly the
dust-side counterpart of the parent problem --- our own N-body
propagation of the grains over the same centuries --- and is therefore
not removable by anchoring the ejection site alone.
We also tested the obvious refinement --- anchoring each trail to the
orbit solution fitted to the \emph{nearest historical apparition} rather
than to the back-integrated modern one.  It does not help.  Repeating
the three epoch-matched hindcasts with per-apparition anchoring leaves
the radial distances essentially unchanged
($\lesssim2\times10^{-4}$~au) but degrades the peak times from 1--40~min
to 2.5--3.4~h.  The reason is that those solutions are fitted to
seventeenth- to nineteenth-century visual astrometry, so between
apparitions they diverge from the modern CCD-based solution by
$0.04$--$0.11$~au --- more than the error they were meant to remove.  We
therefore anchor to the modern solution as propagated by JPL.
\subsection{Uncertainty classes}
\label{sec:conf}
Following the 2019--2025 forecast scorecard assembled from the literature
(timing reliable; amplitude errors of $\times$2--3 for storm-calibrated
trails, $\times$3--5 for well-observed parents without trail calibration,
and 1--2 orders of magnitude for disruption-origin or poorly known
parents; \citealt{egal2020review,egal2023,vida2024}), every forecast is
issued with a confidence class --- A (0.5--2.0), B (0.29--3.5), C (0.10--10) --- and the ZHR profile carries
the corresponding band.  The class is set per shower from that shower's
own hindcast scorecard rather than assumed: the Leonids are issued at
class B --- not because the three anchored storm hindcasts are poor
(they span $\times0.89$--$\times1.25$) but for two reasons the scorecard
alone does not show.  The completeness audit of Sect.~\ref{sec:limits}
finds the same runs generating maxima that were never observed; and the
constant those hindcasts fix is itself uncertain by a factor 1.3--1.5,
because each anchor moves by a factor 1.5--2.2 between ejection
realisations of the same configuration (Sect.~\ref{sec:calib}).  A class
narrower than B would claim a calibration the calibration does not have.
\section{Calibration and validation}
\label{sec:valid}
\label{sec:calib}
\subsection{The 1998--2002 Leonid sequence}
\label{sec:leo9802}
The five consecutive returns of 1998--2002 are the strongest test a
Leonid trail model can be given, and we use them as the primary
validation of everything that follows.  They are not five repetitions of
one measurement: each year probes a different part of the machinery, and
each has a published rate and time to be checked against.
\begin{itemize}
\item \textbf{1998} is the \emph{negative} test for young trails.  No
young trail passed close to Earth, the classical young-trail theory
correctly predicted no storm \citep{mcnaught1999}, and what did appear
was a fireball-rich outburst traced to the resonant 1333 trail, twenty
revolutions old \citep{asher1999res}.  A model that manufactures a young
storm in 1998 is wrong no matter how well it does elsewhere; a model
that produces the observed low young-trail rate has its
$f_{\rm M}$ dilution and coherence cutoff under control.  The year is
scored in Table~\ref{tab:leo9802} but not drawn: with no trail section
on Earth's path there is nothing for a cross-section panel to show.
\item \textbf{1999} is the calibration anchor: a clean single-trail
storm (1899, three revolutions) with a sharply measured maximum,
ZHR $3700\pm100$ at 02:02~UT
($\lambda_\odot = 235.285^\circ$; \citealt{arlt1999}).  It fixes $Z_{\rm storm}$ and
tests the radial kernel at $|r_{\rm E}-r_{\rm D}| \lesssim 10^{-3}$~au.
\item \textbf{2000} is the discriminating case, and the one this model
fails.  Three outbursts were recorded --- ZHR $130\pm20$ on November 17
at 08:07~UT, then $\approx290$ at 03:24 and $\approx480$ at 07:51~UT on
November 18 \citep{arlt2000} --- and all three had been predicted, the
first from the two-revolution 1932 trail \citep{lyytinen2000}, the
others from the eight-revolution 1733 and four-revolution 1866 trails
placed by \citet{asher1999imc} at
$r_{\rm E}-r_{\rm D} \simeq +8\times10^{-4}$~au.  Because none of the
three is a storm, the year tests the cross-section at intermediate
distance in both directions at once: a model can fail here by being too
loud or too quiet, and ours does both.
\item \textbf{2001 and 2002} test \emph{attribution and timing} in the
hardest case: two maxima per night from trails of different ages
(1767 at seven revolutions, 1866 at four), separated by
7--8~h.  The 2002 pair was measured from the air by the Leonid MAC
mission at 04:03 and 10:49~UT \citep{abe2003}.
\end{itemize}
Taken together the sequence constrains the amplitude scale (1999), the
radial fall-off at intermediate distance (2000), the age dilution
(1998), and the along-trail geometry and attribution (2001, 2002).  Table~\ref{tab:leo9802} collects
the comparison, and Figs.~\ref{fig:year1999}--\ref{fig:year2002} show
1999--2002 on exactly the axes used later for the
forecast years, so that hindcast and prediction are read the same way.
All five hindcasts use \emph{epoch-matched} datasets: a separate
$10^{5}$-grain simulation per year, snapshotted at that year's encounter,
so that no result depends on two-body propagation across more than a
fraction of a revolution (Sect.~\ref{sec:systematics}).  All five share
the configuration adopted for the forecasts --- the same ejection law,
the same $\beta \le 3\times10^{-3}$ cut, the same anchoring, the same
node correction.  This matters more than it sounds: an earlier version
of this validation mixed hindcasts built with $\beta \le
1.5\times10^{-3}$ with forecasts built at $3\times10^{-3}$, and the
resulting calibration constant was wrong by a factor of three while
every internal consistency check passed.  The dispersion of the
amplitude ratios is the diagnostic that exposed it: across a mixed
configuration the three storm ratios spanned a factor of five, and on
the self-consistent one they span a factor of 1.4.
\begin{table}[t]
\centering
\footnotesize
\caption{The 1998--2002 Leonid sequence, scored in both directions at once.
A row appears if the model produces a maximum above ZHR~20 --- asking
whether the record contains it --- \emph{or} if the record holds a
documented maximum --- asking whether the model reproduces it.  Rows of
the first kind with no counterpart are marked \textit{not observed};
each is below ZHR 140, within the night-to-night scatter of the annual
shower, and none would have been reported as an outburst.  The
\emph{sum} row compares the model's highest rate of the year with the
highest observed rate of that year, which in 2000 falls on a different
night.  Modelled values are \emph{this} model --- the epoch-matched,
ephemeris-anchored simulation of
Sects.~\ref{sec:stream}--\ref{sec:forecast}; comparisons with the
published models of \citet{mcnaught1999} and \citet{lyytinen2000} are
made separately in Sect.~\ref{sec:t1932} and Table~\ref{tab:lvf}.
$r_{\rm E}-r_{\rm D}$ follows the sign convention of
\citet{mcnaught1999} (positive = trail inside Earth's orbit) and $n$ is
the number of integrated grains supporting the section.  Observed rates
and times follow \citet{arlt1998} for 1998 (peak of the bright-meteor
component, and the ZHR $180\pm20$ storm component), \citet{arlt1999} for
1999, \citet{arlt2000} for 2000,
\citet{arlt2001} for 2001 (Bulletin 17, including the
six-revolution maximum at 11:39~UT), \citet{arlt2002} and
\citet{abe2003} for 2002 and its airborne measurements, \citet{brown2002}
for the 1999 sub-maximum and \citet{jenniskens2006} otherwise; the 1998
fireballs are attributed to the resonant 1333 trail by
\citet{asher1999res}, which lies outside the 140-yr ejection span of the
1998 dataset.}
\label{tab:leo9802}
\begin{tabular}{llrlrrlr}
\toprule
Year & Trail & \multicolumn{2}{c}{This model} & $r_{\rm E}-r_{\rm D}$ &
$n$ & Observed & Ratio \\
\cmidrule(lr){3-4}
 & & ZHR & UT & [$10^{-3}$~au] & & & \\
\midrule
1998 & 1965 & 72 & Nov 17 19:18 & $+6.45$ & 1211 & --- & \textit{not observed} \\
 & 1333 & --- & --- & --- & --- & 340 @ Nov 17 01:40 & \textit{outside span} \\
 & \emph{sum} & 101 & Nov 17 19:18 & --- & --- & 180 @ Nov 17 20:30 & $\times$0.56 \\
1999 & 1932 & 511 & Nov 18 01:38 & $+1.34$ & 899 & +100--300 @ Nov 18 01:43 & $\times$1.7--5.1 \\
 & 1965 & 32 & Nov 18 01:53 & $+4.38$ & 1733 & --- & \textit{not observed} \\
 & 1899 & 3293 & Nov 18 03:08 & $-0.87$ & 449 & 3700 @ Nov 18 02:02 & $\times$0.89 \\
 & \emph{sum} & 3766 & Nov 18 03:08 & --- & --- & 3700 @ Nov 18 02:02 & $\times$1.02 \\
2000 & 1932 & 130 & Nov 17 07:54 & $-1.53$ & 429 & 130 @ Nov 17 08:07 & $\times$1.00 \\
 & 1965 & 342 & Nov 17 08:24 & $+3.01$ & 307 & --- & \textit{not observed} \\
 & 1733 & 7 & Nov 18 07:24 & $+3.02$ & 35 & 290 @ Nov 18 03:24 & $\times$0.02 \\
 & 1866 & 41 & Nov 18 07:39 & $+2.85$ & 63 & 480 @ Nov 18 07:51 & $\times$0.09 \\
 & \emph{sum} & 493 & Nov 17 08:09 & --- & --- & 480 @ Nov 18 07:51 & $\times$1.03 \\
2001 & 1965 & 280 & Nov 17 14:10 & $+2.19$ & 99 & --- & \textit{not observed} \\
 & 1767 & 2032 & Nov 18 10:40 & $+0.29$ & 51 & 1620 @ Nov 18 10:39 & $\times$1.25 \\
 & 1800 & 21 & Nov 18 12:25 & $+2.54$ & 45 & 650 @ Nov 18 11:39 & $\times$0.03 \\
 & 1866 & 602 & Nov 18 17:25 & $+1.65$ & 64 & 3730 @ Nov 18 18:16 & $\times$0.16 \\
 & \emph{sum} & 2073 & Nov 18 10:40 & --- & --- & 3730 @ Nov 18 18:16 & $\times$0.56 \\
2002 & 1767 & 2248 & Nov 19 04:50 & $+0.10$ & 54 & 2510 @ Nov 19 04:10 & $\times$0.90 \\
 & 1800 & 108 & Nov 19 05:35 & $+1.87$ & 55 & --- & \textit{not observed} \\
 & 1866 & 1497 & Nov 19 08:50 & $+1.02$ & 100 & 2940 @ Nov 19 10:47 & $\times$0.51 \\
 & 1733 & 20 & Nov 19 09:05 & $-3.33$ & 81 & --- & \textit{not observed} \\
 & \emph{sum} & 2825 & Nov 19 05:35 & --- & --- & 2940 @ Nov 19 10:47 & $\times$0.96 \\
\bottomrule
\end{tabular}
\end{table}
\subsection{Calibration constant and scorecard}
$Z_{\rm storm}$ is fixed per shower by the documented storm record; for
the Leonids the three anchored, epoch-matched storm hindcasts of
1999--2002 are used jointly.  Requiring the geometric mean of the three
model/observed amplitude ratios to be unity gives
$Z_{\rm storm} = 8.6\times10^{3}$, at which the realised ratios are
$\times0.89$ (1999, 1899), $\times1.25$ (2001, 1767) and $\times0.90$
(2002, 1767), geometric mean $1.00$.  Three independent storms, three
different trail ages and two different centuries of ejection, matched by
one constant to within a factor 1.25 --- that consistency, rather than
the value of the constant, is what the calibration buys.  (The constant
is smaller than the $1.8\times10^{4}$ of earlier versions of this model
only because the population-index conversion of Sect.~\ref{sec:maglim}
now supplies part of the same scaling explicitly.)

How well an anchor is determined is worth measuring rather than
assuming, and the three are not equally determined.  Repeating each
hindcast with independent ejection seeds at fixed configuration --- same
era, same return count, same anchoring, the published grain count ---
gives

\begin{itemize}
\item 1999 (1899, 449 grains): ZHR $4.0\times10^{3}\pm0.8\times10^{3}$
across four realisations, a factor 1.50, with the node fixed at
$r_{\rm E}-r_{\rm D} = -0.84\pm0.07\times10^{-3}$~au;
\item 2002 (1767, 54 grains): ZHR $2.5\times10^{3}\pm0.6\times10^{3}$, a
factor 1.64, node $+0.08\pm0.09\times10^{-3}$~au, all four peaks within
15~min of each other;
\item 2001 (1767, 51 grains): three realisations land within half an
hour of the observed 10:39~UT at ZHR 1126--2442, a factor 2.2 --- and
the fourth misses the encounter altogether, placing the section
$3.6\times10^{-3}$~au ($7\sigma_r$) out for ZHR 21 at 05:25~UT.
\end{itemize}

\noindent A subsampling study of the 1999 trail shows the rate converged
in grain count above $\sim8\times10^{3}$ grains, so what remains is the
burst schedule: with at most 40 ejection bursts per return, which epochs
along the perihelion arc are sampled sets the realisation noise, and a
larger particle count does not reduce it.  Two things follow.  The
geometry is robust where the encounter is a hit and fragile where it is
not --- the two anchors that pass within $1\sigma_r$ hold their nodes to
$0.1\times10^{-3}$~au across realisations, while the 2001 section, met at
the edge of the profile, is lost by one draw in four.  And the
consistency the calibration reports --- three storms matched by one
constant to within a factor 1.25 --- is a property of the geometric mean
of three quantities that individually span factors of 1.5--2.2, not of
three individually precise measurements.  Propagating those spreads
through the geometric mean leaves $Z_{\rm storm}$ itself uncertain by
about a factor 1.3--1.5.  That is the reason to fix it on three storms
jointly rather than on the best-observed one, and the reason the Leonid
forecasts are issued at class~B.
The 2001/2002
reference peaks are the global visual IMO analyses of
\citet{arlt2001} and \citet{arlt2002} throughout, so that every observed
rate in Table~\ref{tab:leo9802} comes from one homogeneously reduced
series; the storm activity record follows \citet{jenniskens2006}.

The 2002 pair was also measured from the air, and the two determinations
are worth setting side by side because they bound the observational
uncertainty this model is calibrated against.  \citet{arlt2002} place
the seven-revolution (1767) maximum at $04^{\rm h}10^{\rm m}\pm1$~min~UT
with ZHR $2510\pm60$ and the four-revolution (1866) maximum at
$10^{\rm h}47^{\rm m}\pm1$~min with ZHR $2940\pm210$, from 528 observing
hours and 57\,045 Leonids.  The Leonid MAC airborne campaign
\citep{abe2003} recorded the same twin maxima at 04:03 and 10:49~UT ---
7~min earlier and 2~min later respectively.  Both determinations agree
to within about a tenth of the 39- and 25-min full widths
\citeauthor{arlt2002} measure for the two peaks, so the choice between
them does not affect any conclusion here; we adopt the IMO times and
rates for consistency with the other four years.  Table~\ref{tab:leo9802}
and Figs.~\ref{fig:year1999}--\ref{fig:year2002} summarize the hindcast
performance.
One further comparison places the amplitude on an external scale.  The
2002 encounter was predicted in advance by six independent groups, and
\citet{arlt2002} tabulated all six against their own measurement;
Table~\ref{tab:pred2002} adds our hindcast to that comparison.  It is
not a fair contest --- ours is a hindcast and theirs were predictions
--- but it is the only place in this paper where the model's amplitude
can be judged against what the field actually achieved on the same
night.  On the seven-revolution 1767 trail our $\times0.90$ is the
closest of the set to the measured rate, the published dynamical
predictions spanning $\times0.40$ to $\times1.79$; on the
four-revolution 1866 trail our $\times0.50$ is the furthest from it.
That asymmetry --- good on the trail this model reproduces, poor on the
one it does not --- is the same signature the completeness audit
isolates in Sect.~\ref{sec:limits}, and it is visible here against six
independent yardsticks rather than against the observation alone.  Our
timing is the weakest entry in the table: every published prediction is
early by 6--22~min and ours is 40~min late.

\begin{table}[t]
\centering
\small
\caption{The 2002 Leonid encounter as a blind benchmark.  All six
entries above the rule were published \emph{before} the event and are
taken from Table 2 of \citet{arlt2002}, which compiled them
against the same IMO analysis used throughout this paper; the last two
are phenomenological rather than dynamical.  Our row is a hindcast, not
a prediction, and is placed last for that reason --- it is not a fair
competitor, and we include it only because it is the one comparison
that puts this model's amplitude on a scale the reader can judge.  On
the seven-revolution 1767 trail our rate is the closest of the set to
the measured $2510\pm60$, and on the four-revolution 1866 trail it is
the furthest, which is the same asymmetry the completeness audit of
Sect.~\ref{sec:limits} finds throughout.  The timing runs the other
way: every published prediction is early by 6--22~min and ours is late
by 40, the largest error in the column.  Two of the six predictions
appeared only online and are cited here through
\citeauthor{arlt2002}'s compilation.}
\label{tab:pred2002}
\begin{tabular}{lllll}
\toprule
Source & \multicolumn{2}{c}{1767 trail (7 rev)} &
\multicolumn{2}{c}{1866 trail (4 rev)} \\
\cmidrule(lr){2-3}\cmidrule(lr){4-5}
 & UT & ZHR & UT & ZHR \\
\midrule
\citet{lyytinen2000}          & 04:02 & 4500 & 10:44 & 7400 \\
Lyytinen et al.\ (2002)$^{a}$ & 04:03 & 3500 & 10:40 & 2600 \\
\citet{mcnaught2002}          & 03:56 & 1000 (816--2000) & 10:34 & 6000 (2900--6000) \\
Vaubaillon (2002)$^{a}$       & 04:04 & 3600 & 10:47 & 3200 \\
Jenniskens (2002)$^{a,b}$     & 03:48 & 5900 & 10:23 & 5400 \\
\citet{langbroek2002}$^{b}$   & --- & 2000--5700 & --- & 2400--5200 \\
\midrule
\textbf{Observed} \citep{arlt2002} & \textbf{04:10} & $\mathbf{2510\pm60}$
                                   & \textbf{10:47} & $\mathbf{2940\pm210}$ \\
\midrule
This model (hindcast)         & 04:50 & 2209 & 09:05 & 1458 \\
\bottomrule
\end{tabular}
\\[2pt]
\begin{minipage}{0.92\linewidth}\footnotesize
$^{a}$ published online only; quoted here from Table 2 of
\citet{arlt2002}.
$^{b}$ phenomenological, not a dust-trail integration.
\end{minipage}
\end{table}

The decisive test of the anchoring is not the amplitude but the
\emph{attribution}.  Before anchoring, the epoch-matched 2001 and 2002
runs placed the dominant crest in the 1833 trail, at odds with the
classical analyses which assign both storms to the 1767 and 1866 trails
\citep{mcnaught1999,asher1999,lyytinen2001}; this discrepancy was the
principal open problem of the earlier version of this model.  With the
ejection sites anchored, the dominant modelled trail in both years is
the 1767 trail, and its predicted maximum falls 1~min after the observed
2001 peak (10:40 vs.\ 10:39~UT) and 40~min after the observed 2002 peak
(04:50 vs.\ 04:10~UT).  The corresponding radial miss distances become
$+2.9\times10^{-4}$ and $+1.0\times10^{-4}$~au, against
$+3.5\times10^{-3}$~au before anchoring --- the age-proportional outward
bias is gone.  The model still fails to reproduce the second,
1866-trail maximum of either year (18:16~UT in 2001, 10:47~UT in 2002),
reaching only $\times0.16$ and $\times0.51$ of the observed rates and
placing that trail $+1.7\times10^{-3}$ and $+1.0\times10^{-3}$~au inside
Earth's orbit; we return to this in Sect.~\ref{sec:discussion}.
The 1999 anchor itself is reproduced at $\times0.89$ with the peak
1.1~h late, and the 1899-trail node at
$r_{\rm E}-r_{\rm D} = -8.7\times10^{-4}$~au, i.e.\ the trail centre
just outside Earth's orbit --- the same side, and the same order of
magnitude, as in Fig.~8 of \citet{asher1999imc}.
\subsection{Completeness: every maximum the model produces}
\label{sec:limits}
A hindcast table built only from the trails the classical analyses
nominate can measure how well the model reproduces events that are known
to have occurred, and nothing else.  It is silent on the complementary
and, for a forecasting model, equally consequential question: what else
does the model predict, and does the record contain it?
Table~\ref{tab:leo9802} is therefore scored in both directions --- every
model maximum above ZHR 20, and every documented maximum, whether or not
the model produces one --- so that misses and excesses are counted on the
same page.  The audit is not flattering.  Three findings stand out.
First, the reproduced storms survive the wider view intact: in 2001 and
2002 the 1767 maxima remain the model's dominant features at the correct
times.  The wider view is not free of cost, however: the additional
1965 entries have no counterpart in the record, and the 1800 section of
2001 does --- \citet{arlt2001} resolve a six-revolution maximum of
ZHR $650\pm40$ at 11:39~UT, 46~min from where the model puts that trail,
and the model gives it ZHR 21, a factor 31 short.  That miss is invisible
to a table built from the nominated trails alone.
A second miss of the same kind falls outside the table because it cannot
be attributed to a trail at all: \citet{arlt1999} report a clear second
1999 outburst at $\lambda_\odot = 235.87^\circ \pm 0.04^\circ$
(November 18, $16^{\rm h}\pm15^{\rm m}$~UT) reaching ZHR $180\pm20$,
some fourteen hours after the storm.  Our model produces only ZHR 19
there, a factor nine short, and assigns it to no section: no trail in
the epoch-matched 1999 dataset --- four trails spanning the 140~yr to
1998 --- has a node within $0.35^\circ$ of that longitude.  The most
likely explanation is the one already established for the 1998
fireballs, namely a trail older than the ejection span the dataset
carries \citep{asher1999res}; testing it would require the
eighteen-return configuration used for the forecasts, run epoch-matched
to 1999.  We record the miss here rather than in
Table~\ref{tab:leo9802}, whose rows are organised by trail, but it
belongs to the same account: of the documented 1998--2002 maxima, the
ones this model misses outright are the old, weak and unattributed ones,
not the storms.
Second --- and this is what the audit was for --- the two-revolution
1932 trail was grossly over-weighted in both 1999 and 2000.  The table
already carries the empirical density factor derived in
Sect.~\ref{sec:t1932}; \emph{without} it the physical model places that
section $1.3\times10^{-3}$~au inside Earth's orbit and assigns it ZHR
$9.8\times10^{3}$ at 01:38~UT in 1999 --- half an hour before, and nearly
three times as strong as, the genuine 1899-trail storm it reproduces
correctly --- taking the summed profile to ZHR $1.2\times10^{4}$, a
factor 3.3 above the observed 3700 even though the trail the storm is attributed to
is matched to $\times0.89$.  In 2000 the same section gives ZHR $2.4\times10^{3}$ on
November 17 against a measured $130\pm20$.  Neither encounter is
fictitious: both were predicted independently and both were detected, the
1999 one as a sub-maximum inside the storm profile and the 2000 one as an
outburst in its own right.  What the audit exposes is therefore not
invention but an eighteenfold error of \emph{rate} on one trail --- and a
table restricted to the nominated trails would have hidden it, because
1999 would have read $\times0.89$ and nothing else.
Third, the sections carrying the largest excesses share a signature: an
absolute density factor far above what their age allows.  The 1965
section in 1998, one revolution old, reaches $f_{\rm M} = 1.93$ and the
1932 section in 1999, two revolutions old, $1.28$ --- both above unity ---
while the same 1932 section in 2000 reaches $0.92$
(Table~\ref{tab:lvf}), where two revolutions of unperturbed linear
stretching should give $\simeq0.5$.  Values above unity are unphysical by
construction: $f_{\rm M}$ is
normalised so that a section which has undergone exactly one revolution
of unperturbed linear stretching scores 1, and a two-revolution section
should score $\simeq 0.5$.  What the estimator is measuring instead is
the clumpiness of a trail that has not yet stretched into a smooth tube,
where the along-trail density kernel resolves individual condensations
rather than a mean linear density.  The model is therefore quantitatively trustworthy over
the three- to eight-revolution band that both the calibration storms and
the 2030s forecasts occupy, and should not be used below three
revolutions without recalibration.  The revolution count is not the
whole criterion, however.  Sect.~\ref{sec:leo2009} shows the same
inflation on a \emph{seven}-revolution section met $8\sigma_r$ out in
its wing, and establishes the sharper test: $f_{\rm M}$ tracks the
analytic $1/n_{\rm rev}$ to a median ratio $0.95$ for encounters within
$3\sigma_r$ of the trail core and is unconstrained beyond $5\sigma_r$.
Every encounter in Table~\ref{tab:leo9802} that the model
over-predicts, and both sections carrying the 2034 forecast --- the 1932
trail met at $1.1\sigma_r$ and the 1733 trail at $2.8\sigma_r$ ---
should be read against that test as well as against $n_{\rm rev}$.  We flag this explicitly rather than
tune it away: with only two young-trail encounters in the record, any
correction fitted here would be fitted to two points.
\subsection{The 1932 trail: right trail, right time, wrong rate}
\label{sec:t1932}

The 1932 section carries both large excesses in Table~\ref{tab:leo9802},
and it also dominates our 2034 forecast, so we examined it against the
literature and in detail.

\paragraph{It is not a spurious trail}
Our first reading of Table~\ref{tab:leo9802} was that the model invents
two encounters.  It does not.  Both were predicted independently, before
the fact, and both have observational counterparts.  For 1999,
\citet{mcnaught1999b} tabulated a 1932-trail encounter with maximum
expected at 01:44~UT; \citet{arlt1999} concluded from numerical modelling
that \emph{both} the 1899 and 1932 ejecta contributed to the storm; and
\citet{brown2002} found a significant feature in the measured flux curve
at $\lambda_\odot = 235.272^\circ$ ``corresponding to the expected time
for encounter with material released in 1932'', matching a visual
enhancement at 01:40--01:43~UT and a video plateau at 01:39--01:53~UT
\citep{rendtel2000}, while cautioning that a statistical fluctuation
cannot be excluded.  For 2000, \citet{lyytinen2000} predicted a
two-revolution (1932) encounter at 07:50~UT on November 17 with
ZHR 215, and the IMO global analysis recorded a maximum of
ZHR $130\pm20$ at 08:07~UT that night \citep{arlt2000}.  Our model places
these at 01:38~UT and 07:39~UT --- 6 and 11~min from the published
predictions, and 28~min before the 2000 measurement.  The trail
identification and the timing are right.  What is wrong is the rate:
$\times18.6$ of the observed 2000 outburst, and a dominant storm rather
than a sub-feature in 1999.

\paragraph{It is not sampling noise}
Subsampling the $10^{5}$-grain 1999 dataset by factors 2, 4 and 8 leaves
the section's concentration factor $f_{\rm M} n_{\rm rev}$ at
$2.77$, $2.79$, $2.84$, $2.76$ and its node at
$r_{\rm E}-r_{\rm D} = +1.30$ to $+1.34\times10^{-3}$~au.  Neither
quantity moves outside 3\% over an eightfold change in grain count, so
raising the particle number cannot remove the excess.  The concentration
is also not an artefact of the kernel density estimator: a plain
histogram of the grains' node-arrival times, with no smoothing, gives an
independent concentration of $2.20$.  Nor is it an artefact of how far
back the dataset reaches: an entirely separate $2.4\times10^{5}$-grain
run extending to eight returns and a 250-yr back-integration returns the
same section at $r_{\rm E}-r_{\rm D} = +1.23\times10^{-3}$~au with a
concentration of $2.85$ (again flat, $2.65$--$2.90$, under eightfold
subsampling).  In that run the \emph{reproduced} 1899 storm shifts by
less than the excess does: its node moves to $-1.07\times10^{-3}$~au and
its amplitude to $\times1.32$ of the observed 3700, against
$\times0.89$ in the calibration set, while the 1932 section stays where
it was in both node and concentration.  More grains and a longer parent
history move what the model gets right by a factor of order the
calibration spread, and leave what it gets wrong exactly where it was.

\paragraph{It is not the radial profile}
We rescanned the cross-section shape against the whole 1998--2002 record,
re-deriving $Z_{\rm storm}$ for every configuration so that the scan
measures shape rather than scale.  Over generalized-Lorentzian exponents
$p = 1.35$--$3.5$, a Gaussian $f_r$ as used by \citet{mcnaught1999}, and
$\sigma_r = 4$--$6\times10^{-4}$~au, the 1999 summed profile never falls
below $\times1.23$ of the observed rate, and the configurations that
suppress it most ($p = 3.5$) destroy 1998 and 2000 instead, tripling the
RMS error over the five yearly totals.  The adopted
$(p, \sigma_r) = (1.35, 5\times10^{-4}\,{\rm au})$ sits at the optimum
(RMS 0.20 dex over the five yearly totals).

\paragraph{It is the mean-anomaly factor, and only for young trails}
Table~\ref{tab:lvf} compares our nodes and $f_{\rm M}$ values with
Table~I of \citet{lyytinen2000}, the only published tabulation that gives
both quantities per trail.  The comparison is unusually clean.  For the
seven-revolution 1767 trail --- the two encounters this model reproduces
best --- our $f_{\rm M}$ agrees with theirs to 1\% (2001) and 11\%
(2002), and our node to $2$--$7\times10^{-4}$~au.  Across the ten
entries with $n_{\rm rev} \ge 4$ our $f_{\rm M}$ has median ratio
$0.61$: acceptable, and biased low if anything.  The two-revolution 1932
trail is the \emph{only} entry we place above theirs, at $\times1.77$
($f_{\rm M} = 0.97$ against $0.55$), even though we agree with them on
its node to $3.3\times10^{-4}$~au and on the side of Earth's orbit it
passes.  The same estimator that is right at seven revolutions is
$1.8$--$2.7$ times too generous at two.

The remaining factor of about four is one \citeauthor{lyytinen2000}
identify in their own table: the comment against that row is
\emph{``mostly faint''}.  The 2000 encounter meets the 1932 trail at
$\Delta a_0 = 0.30$~au, far out on the ejection distribution, and the
grains that reach Earth at that offset do so through large
radiation-pressure period changes --- which selects small particles.  A
section rich in $0.1$--$0.5$~mm grains delivers radar and telescopic
meteors but few of visual magnitude, and \citeauthor{lyytinen2000}
accordingly still over-predicted the visual rate by $\times1.65$.  Our
model assigns every trail the same ejection size distribution and the
same $\beta \le 3\times10^{-3}$ cut, so it reproduces the section's
\emph{position} but not its depletion in visually detectable grains.
The product $1.8 \times 4$ accounts for a factor $\simeq7$, which is the
excess the \emph{physical} model alone produces.

The adopted configuration is a further factor 2.9 above that, and the
cause is the population-index conversion of Sect.~\ref{sec:maglim}
behaving exactly as it should.  Being genuinely faint-enriched, the 1932
section is measured at the slope ceiling ($u = 3.61$ raw, clipped to
$3.5$, $r = 2.15$) against $r = 1.33$--$1.47$ for the three calibrating
storm sections, so its extrapolation factor $r^{6.5-4.2}$ is $5.89$ where
theirs average $2.23$.  Since $Z_{\rm storm}$ is fixed on those storms it
absorbs the $2.23$, and the 1932 section retains the remaining
$\times2.6$.  The full budget in the adopted configuration is therefore
$1.8 \times 4 \times 2.6 = 18.7$, against the $\times18.6$ measured ---
and it is worth being precise about which of the three terms is an
error.  The conversion is not one: a steeper size distribution genuinely
does imply more meteors between $+4.2$ and $+6.5$.  What it cannot
represent is the deficiency \citeauthor{lyytinen2000} point at, which is
in $N$ --- the number of visually detectable meteoroids at that node ---
and not in the exponent applied to it.  Multiplying an $N$ that is
already too large by a correctly larger $r^{6.5-4.2}$ compounds the
error instead of curing it.  The extrapolation also assumes a
sub-$0.478$~mm population at the same node, which the size-range
experiment below shows is precisely what a trail does \emph{not} have.

\paragraph{Widening the size range does not fix it}
\citeauthor{lyytinen2000}'s ``mostly faint'' remark suggests an obvious
remedy: our size sampler is truncated at $a \ge 0.478$~mm by the
$\beta \le 3\times10^{-3}$ cut, well above the $0.159$~mm that a $+6.5$
magnitude Leonid requires, so the magnitude cut in the forecast never
fires --- 100.0\% of grains clear it in every section of every year
examined here --- and the model has no faint population whose depletion
it could represent.  We therefore repeated the 2000 encounter with the
sampler extended down to the visual limit itself
($3.2\times10^{5}$ grains, eight returns, $a \ge 0.159$~mm, a sixteenfold
increase in grain number at fixed mass budget).  To isolate the effect
of the size range itself, the ratios in this paragraph are quoted
\emph{without} the population-index conversion, i.e.\ against the
$\times7$ baseline of the physical model rather than the $\times18.6$ of
the adopted configuration.

For the 1932 trail the result is what the hypothesis predicts.  Its
mean-anomaly factor falls from $f_{\rm M} = 0.97$ to $0.59$, against
\citeauthor{lyytinen2000}'s $0.55$ --- agreement to 8\% where there had
been a factor 1.8 --- and its amplitude falls from $\times7.0$ to
$\times1.6$ of the observed ZHR $130\pm20$, which is
\citeauthor{lyytinen2000}'s own accuracy on that encounter
($\times1.65$).  Taken alone this looks like the fix.

It is not, for two reasons.  First, the amplitude improvement is only
about 40\% due to $f_{\rm M}$; the rest comes from the section's node
moving \emph{away} from Earth, from $-1.53$ to
$-2.25\times10^{-3}$~au --- that is, further from
\citeauthor{lyytinen2000}'s $-1.20\times10^{-3}$~au than before.  The
rate improved partly because two errors moved in compensating
directions.  Second, and decisively, every other section degrades.  The
one-revolution 1965 trail moves from $+3.01$ to
$+1.18\times10^{-3}$~au with $f_{\rm M}$ rising $0.39 \to 1.96$, and
produces ZHR $2.2\times10^{4}$ on November 17 --- a storm forty-five
times the largest rate recorded that year.  The 1866 node swings from
$+2.85$ to $-7.47\times10^{-3}$~au, changing side; the 1767 node moves
from $+3.96\times10^{-3}$ to $+1.02\times10^{-2}$~au.  The summed 2000
profile goes from $\times2.2$ of the observed maximum to $\times46$.

The reason is structural.  Radiation pressure displaces a grain's node in
proportion to $\beta$, so a trail sampled over $\beta \le 3\times10^{-3}$
is a narrow tube, while one sampled to the visual limit
($\beta \le 9\times10^{-3}$) is a \emph{fan} whose width exceeds the
encounter kernel.  Because the added small grains outnumber the original
population sixteenfold, the median node follows them, and a single node
plus a single $f_{\rm M}$ per trail --- the abstraction this model and
\citet{mcnaught1999} both use --- ceases to describe the section at all.
Representing the faint population therefore requires binning each trail
by $\beta$ and treating each bin as its own section with its own node,
cross-section and visible-grain count; it cannot be had by widening a
sampling bound.  We record this as a negative result and retain
$\beta \le 3\times10^{-3}$.  The experiment is also expensive: 12.7~h of
CPU for a single epoch-year, against 7~h for the standard configuration
at three quarters the grain count.

\paragraph{Node distance and the steepness of the profile}
One further sensitivity governs how every near-miss prediction in this
paper should be read.  Figure~8 of \citet{asher1999imc} plots the 1999
cross-section; measured against its own date scale (Nov 17.0 to Nov 18.0
spans one day, $1.72\times10^{-2}$~au) it places the 1899 section
$1.0\times10^{-3}$~au outside Earth's orbit and the 1932 section
$1.9\times10^{-3}$~au inside it.  Our model agrees on the 1899 section to
$0.2\times10^{-3}$~au ($-0.87$ in the calibration set, $-1.07$ in the
eight-return run) but places the 1932 section at $+1.2$ to
$+1.3\times10^{-3}$~au, some $0.6\times10^{-3}$~au too close.  At
$|r_{\rm E}-r_{\rm D}| \simeq 3\sigma_r$ the profile falls by a factor
2.2 per $0.5\times10^{-3}$~au, so a node error that is invisible for a
direct hit becomes a factor of several for a near miss.  \emph{The
predicted rate of a near-miss encounter is exponentially sensitive to the
node distance; that of a direct hit is not.}  This is also why the four-
to nine-revolution trails of Table~\ref{tab:lvf} are under-predicted
despite acceptable $f_{\rm M}$: their nodes sit $1$--$4\times10^{-3}$~au
too far inside Earth's orbit, on the steep flank.

\begin{table}[t]
\centering
\small
\caption{This model against Table~I of \citet{lyytinen2000}, the only
published tabulation giving both the radial miss distance and the
mean-anomaly factor per trail for 2000--2002.  Distances are
$r_{\rm E}-r_{\rm D}$ in $10^{-3}$~au (positive = trail inside Earth's
orbit).  The seven-revolution 1767 trail --- the two encounters this
model reproduces best --- matches their $f_{\rm M}$ to 1\% and 11\%.
The two-revolution 1932 trail, the source of the 2000 over-prediction,
is the only entry whose $f_{\rm M}$ we place \emph{above} theirs.  The
four- to nine-revolution trails have acceptable $f_{\rm M}$ but nodes
displaced $1$--$4\times10^{-3}$~au too far inside Earth's orbit, which
is why they are under-predicted (Sect.~\ref{sec:t1932}).}
\label{tab:lvf}
\begin{tabular}{lll rr rr rr}
\toprule
 & & & \multicolumn{2}{c}{\citet{lyytinen2000}} &
\multicolumn{2}{c}{This model} & \multicolumn{2}{c}{Difference} \\
\cmidrule(lr){4-5} \cmidrule(lr){6-7} \cmidrule(lr){8-9}
Year & $n_{\rm rev}$ & Trail & $r_{\rm E}-r_{\rm D}$ & $f_{\rm M}$ &
$r_{\rm E}-r_{\rm D}$ & $f_{\rm M}$ & $\Delta r$ & $f_{\rm M}$ ratio \\
\midrule
2000 & 2 & 1932 & $-1.20$ & 0.550 & $-1.53$ & 0.922 & $-0.33$ & 1.68 \\
2000 & 4 & 1866 & $+0.80$ & 0.135 & $+2.85$ & 0.098 & $+2.05$ & 0.73 \\
2000 & 8 & 1733 & $+0.80$ & 0.250 & $+3.02$ & 0.132 & $+2.22$ & 0.53 \\
2001 & 4 & 1866 & $+0.25$ & 0.135 & $+1.65$ & 0.094 & $+1.40$ & 0.70 \\
2001 & 5 & 1833 & $+1.78$ & 0.114 & $+2.74$ & 0.010 & $+0.96$ & 0.09 \\
2001 & 6 & 1800 & $+1.35$ & 0.123 & $+2.54$ & 0.058 & $+1.19$ & 0.47 \\
2001 & 7 & 1767 & $-0.43$ & 0.140 & $+0.29$ & 0.136 & $+0.72$ & 0.97 \\
2001 & 9 & 1699 & $+0.10$ & 0.260 & $+3.74$ & 0.123 & $+3.64$ & 0.47 \\
2002 & 4 & 1866 & $-0.04$ & 0.148 & $+1.02$ & 0.164 & $+1.06$ & 1.10 \\
2002 & 5 & 1833 & $+1.48$ & 0.115 & $+2.01$ & 0.020 & $+0.53$ & 0.17 \\
2002 & 7 & 1767 & $-0.13$ & 0.130 & $+0.10$ & 0.114 & $+0.23$ & 0.87 \\
\bottomrule
\end{tabular}
\end{table}

\paragraph{Calibrating the 1932 trail on the 2000 outburst}
The excess is the same size at every 1932 encounter the record
constrains, which invites an empirical fix.  We fix a single density
factor on the 2000 November 17 outburst alone --- the only 1932 encounter
with a measured rate --- and then test it, unfitted, against the other
two.  The factor is $C = 1/18.6$ in the adopted configuration.  It was
$1/20.5$ before the population-index slope was clipped at the ejected
value (Sect.~\ref{sec:maglim}); the 1932 sections are floor-bound in
the sense described there, so part of the excess might have been that
artefact, and re-deriving the factor shows it was not --- 10\% of it,
not a factor.

It survives both tests.  Carried to 1999, the 1932 section falls from
ZHR $9.5\times10^{3}$ to 511, the year's summed profile from
$\times3.3$ to $\times1.02$ of the observed 3700, and --- the part that
matters more than the amplitude --- the 1932 contribution stops being the
dominant peak and becomes a sub-maximum at 01:38~UT riding on the
1899-trail storm at 03:08.  That is precisely the structure
\citet{brown2002} measure: a significant feature in the flux curve at
$\lambda_\odot = 235.272^\circ$, half an hour before the maximum, with
the storm itself from the 1899 trail.  Carried to 1998, the year's total
moves from $\times0.68$ to $\times0.56$ of the ZHR $180\pm20$ storm
component \citep{arlt1998}.  One constant, fixed on one
encounter, corrects three.

The factor is also insensitive to the population-index conversion of
Sect.~\ref{sec:maglim}, and now exactly so: derived without it,
$C = 1/3.1$; derived with it, $C = 1/18.6$; the 2034 maximum is ZHR 1376
either way.  $C$ is fitted and applied on the same trail, and since the
slope ceiling puts both the 2000 and the 2034 section at the same $r$,
whatever the conversion does to one it does to the other and the constant
absorbs it exactly.

\paragraph{Read as a production history}
There is a second reading of that constant, and it is the one
\citet{watanabe2008} develop: if the modelled rate is
$\propto \sum f_{\rm M}$, then the residual between observed and
modelled rates measures the parent's dust production at the returns
that made the trails, so a shower becomes a fossil record of its
comet.  They demonstrate it on the October Draconids, where
normalising on the 1998 outburst under-predicts the 1933 and 1946
storms; correcting each trail's $f_{\rm M}$ by a production factor
$Q(H_{10})$ taken from 21P's archived absolute magnitude --- the comet
was anomalously faint at the 1926 apparition that made the 1998 trail
--- brings all three into agreement.  Our per-trail density factor is
formally that correction, and under this reading 55P delivered some
twenty times less visually detectable dust in 1932 than the uniform
production law of Sect.~\ref{sec:ejection} assumes.

We do not adopt that reading, and the reason is
\citeauthor{watanabe2008}'s own.  Their method requires two conditions:
that the outburst be identified with a trail, and that \emph{the parent
was observed as an active comet at the epoch that formed it}.  They
state explicitly that the Leonids fail the second and exclude them,
because 55P ``was first seen briefly in 1366, and well observed only in
1865 and had its next apparition in 1965''.  The 1932 return is the
extreme case: the comet was searched for and not found, so there is no
$H_{10}$ from which a production factor could be built.  A correction
of this form can still be \emph{measured}, as we measure ours on the
2000 outburst, but it cannot be independently derived, and it therefore
carries no evidence about the comet.  The 1899 trail makes the same
point from the other side: it comes from the other unobserved return
and needs no correction at all, reproducing the 1999 storm at
$\times0.89$.  Meanwhile the decomposition of the preceding paragraphs
--- $f_{\rm M}$, the far-tail size sorting, the population-index
conversion --- accounts for the whole factor without invoking the comet.

The distinction is not academic, because 2034 tests it.  A production
deficit is a property of the entire trail and transfers to any crossing
of it; the far-tail size sorting is a property of the section met in
2000 and does not transfer to the core crossing of 2034.  Applying $C$
there is thus the conservative choice under the production reading and
an extrapolation under ours, and the two give ZHR $1.0\times10^{3}$
against $1.9\times10^{4}$ for the 03:00~UT maximum.  The night
discriminates not only between models but between two readings of the
same empirical constant.

We adopt the correction, and state its limit plainly.  All three
constrained encounters are two-revolution sections met far out on the
ejection distribution ($\Delta a_0 = -0.29$ to $-0.38$~au); the 2034 encounter is a
three-revolution section met essentially at the trail core
($\Delta a_0 = -0.02$~au).  Applying $C$ there assumes the excess is a
property of the trail rather than of its far tail.  The three encounters
share the former reading --- one factor fits sections at three different
node distances and two different sides of Earth's orbit --- but all three
lie in the same regime, so the assumption is an extrapolation and cannot
be tested until 2033 or 2034 itself tests it.

\paragraph{Consequences for 2034}
The stronger of the two 2034 maxima comes from the same 1932 trail, one
revolution older, so the diagnosis above bears directly on it.  Two things
are reassuring and one is not.  First, the encounter is internally
converged: subsampling the section from $5\times10^{3}$ to
$2.5\times10^{4}$ grains moves its node by $6\times10^{-5}$~au
($+0.54$ to $+0.60\times10^{-3}$~au), its concentration by 12\%
($f_{\rm M} = 0.71$--$0.80$) and its rate by 20\%, with no trend in the
last three steps.  Second, and
more usefully, 2034 does not repeat the geometry that makes 2000 hardest.
Measuring the semi-major-axis offset of the grains that actually reach the
node, our model puts the 2000 encounter at $\Delta a_0 = -0.29$~au ---
within 2\% of the $0.30$~au \citet{lyytinen2000} quote for it, an
independent check that we are sampling the same part of the trail they are
--- whereas in 2034 Earth crosses the 1932 trail essentially at its core,
$\Delta a_0 = -0.02$~au.  The ``mostly faint'' depletion that costs a
factor of several in 2000 is a property of that far-tail section and does
not obviously transfer to a core crossing.

What does transfer is the density excess itself, and the empirical factor
above carries it: the 2034 1932 maximum becomes ZHR $1.4\times10^{3}$
rather than $2.6\times10^{4}$.  With the correction applied the night
carries two maxima of comparable height --- the 1932 trail at 03:30~UT and
the nine-revolution 1733 trail at 23:45~UT, ZHR $1.2\times10^{3}$ --- and
without it the early one outweighs the later twentyfold.  Neither the
times nor the node distances depend on the correction, so the night
measures it directly.  What the earlier, frozen-element version of this
forecast made the storm of that night, the eight-revolution 1767 trail at
ZHR $3.5\times10^{3}$, is not part of the picture at all once each year is
computed at its own epoch (Sect.~\ref{sec:systematics}): that section lies
$4.3\sigma_r$ inside Earth's orbit and reaches ZHR 30.

Quoting these numbers in \citeauthor{asher1999imc}'s sign convention
also exposes a residual that the storm hindcasts alone hide.  For the
2000 encounters \citet{asher1999imc} states
$r_{\rm E}-r_{\rm D} \simeq +8\times10^{-4}$~au for both the
eight-revolution (1733) and four-revolution (1866) trails; our anchored,
epoch-matched model places them at $+3.0\times10^{-3}$ and
$+2.9\times10^{-3}$~au --- the correct sign, but three to four times too
far inside Earth's orbit.  The same offset appears in the storm years:
the 1866 trail sits at $+1.7\times10^{-3}$ (2001) and
$+1.0\times10^{-3}$~au (2002) where the observed storms require
$\lesssim10^{-3}$~au, whereas the 1767 trail lands at
$+2.9\times10^{-4}$ and $+1.0\times10^{-4}$~au and the 1899 trail at
$-8.7\times10^{-4}$~au in 1999.  The offset is therefore not a global
bias --- it is confined to particular old trails, and it is what
prevents the model from reproducing the second maximum of 2001 and
2002.  It is not a radiation-pressure artefact either: binning each
trail's grains by $\beta$ moves the node by only
$(0.5$--$0.9)\times10^{-3}$~au across the sampled range, an order of
magnitude too little and in the wrong sense to account for it.  With the ejection
sites now fixed to the observed comet, the remaining candidate is the
century-scale N-body propagation of the grains themselves.
A 12-member Monte
Carlo propagating the full 8-dimensional 55P orbit-solution covariance
of Table~\ref{tab:cov} (including $A_1$, $A_2$) plus an inter-apparition non-gravitational
model error (lognormal, $\sigma_{A_1}=50\%$, $\sigma_{A_2}=30\%$)
through the complete trail simulation, performed with the internal
back-integration at $3.2\times10^{4}$ grains per member, places the 1999
encounter at $\lambda_\odot = 235.295^\circ \pm 0.004^\circ$, i.e.\
13~min later than the observed storm longitude.  Repeating the nominal
orbit with independent ejection samples scatters the same estimator by
$0.0035^\circ$ --- indistinguishable from the ensemble spread --- so
even at this grain count the orbit solution's formal uncertainty
contributes no more to the predicted time than the Monte-Carlo sampling
of the ejection does, and the quoted spread remains an upper limit.
Radially the covariance moves the node by only $1.3\times10^{-4}$~au
($0.26\sigma_r$), against the $10^{-3}$--$10^{-2}$~au and multi-hour
systematic that anchoring removes: two orders of magnitude.  What the
covariance does move is the \emph{rate}, which spans a factor 2.1 across
the ensemble where the ejection controls span only 1.3 --- the radial
profile sensitivity of Sect.~\ref{sec:t1932} seen from the other side,
and a reminder that the amplitude of a near-miss encounter is the
quantity an orbit uncertainty reaches first.  The experiment is
reproducible from the archived covariance matrix of
Table~\ref{tab:cov}.

\begin{figure}[p]
\centering
\includegraphics[height=0.74\textheight]{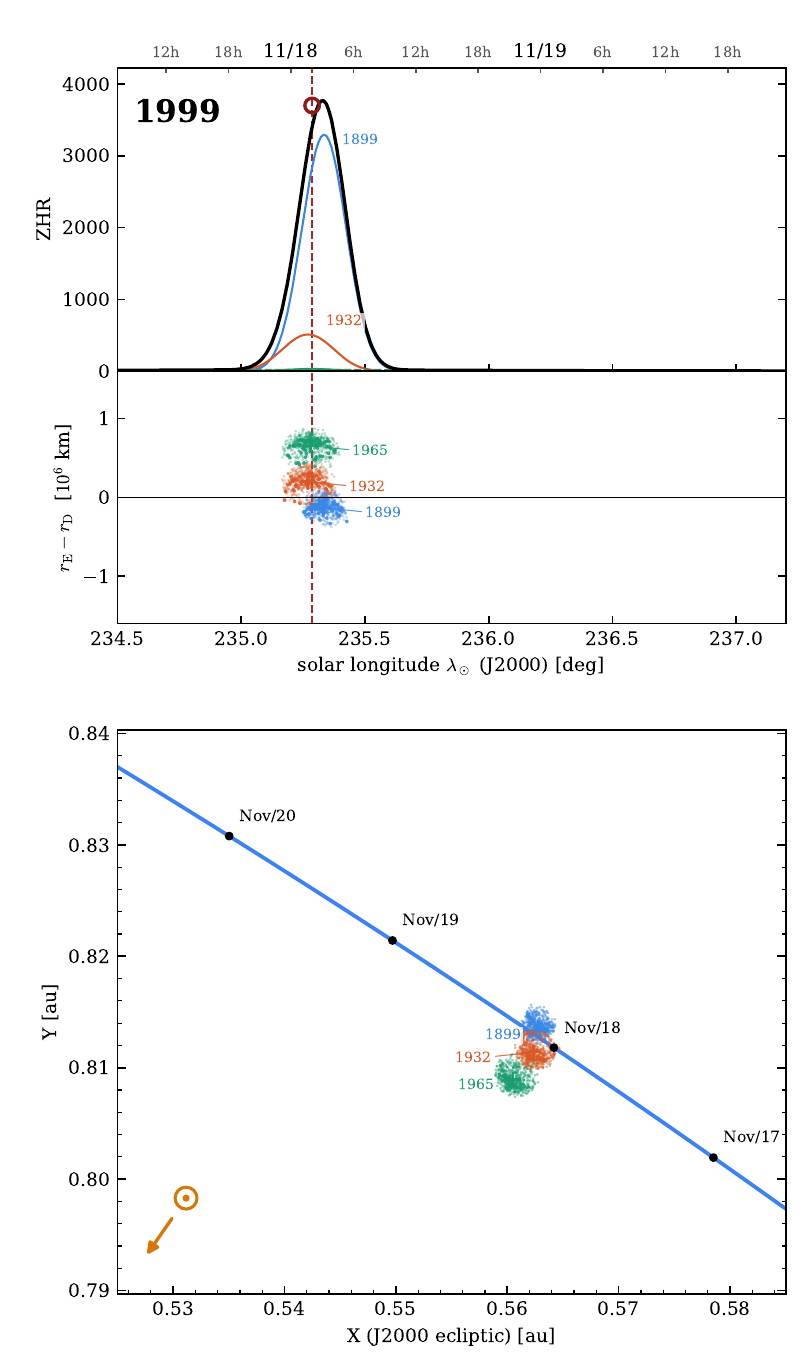}
\caption{Epoch-matched hindcast of the 1999 Leonid encounter, and the
reference for the panels, symbols and conventions of every later year.
\emph{Top}: ZHR on the 15-min grid (dark: total; coloured:
per-trail; green dashed: background), with the observed maxima as circles
and red rules \citep{arlt1999,abe2003}.  \emph{Middle}: nodal
cross-section, $r_{\rm E}-r_{\rm D}$ in the \citet{mcnaught1999}
convention (positive = inside Earth's orbit).  The scatter and its
$1\sigma$/$2\sigma$ ellipses give the grain spread; the cross marks the
encounter the forecast uses, directly below its maximum above.  Solid
points arrive within $T_{\rm lim} = 7$~d of Earth \citep[][10--20\% of a
section]{gockel2000}, and sections wider than $5\sigma_r$ are drawn as
the cross alone.  \emph{Bottom}: ecliptic plane, equal scale, blue =
Earth's orbit, arrow = sunward.
}
\label{fig:year1999}
\end{figure}

\begin{figure}[p]
\centering
\includegraphics[height=0.74\textheight]{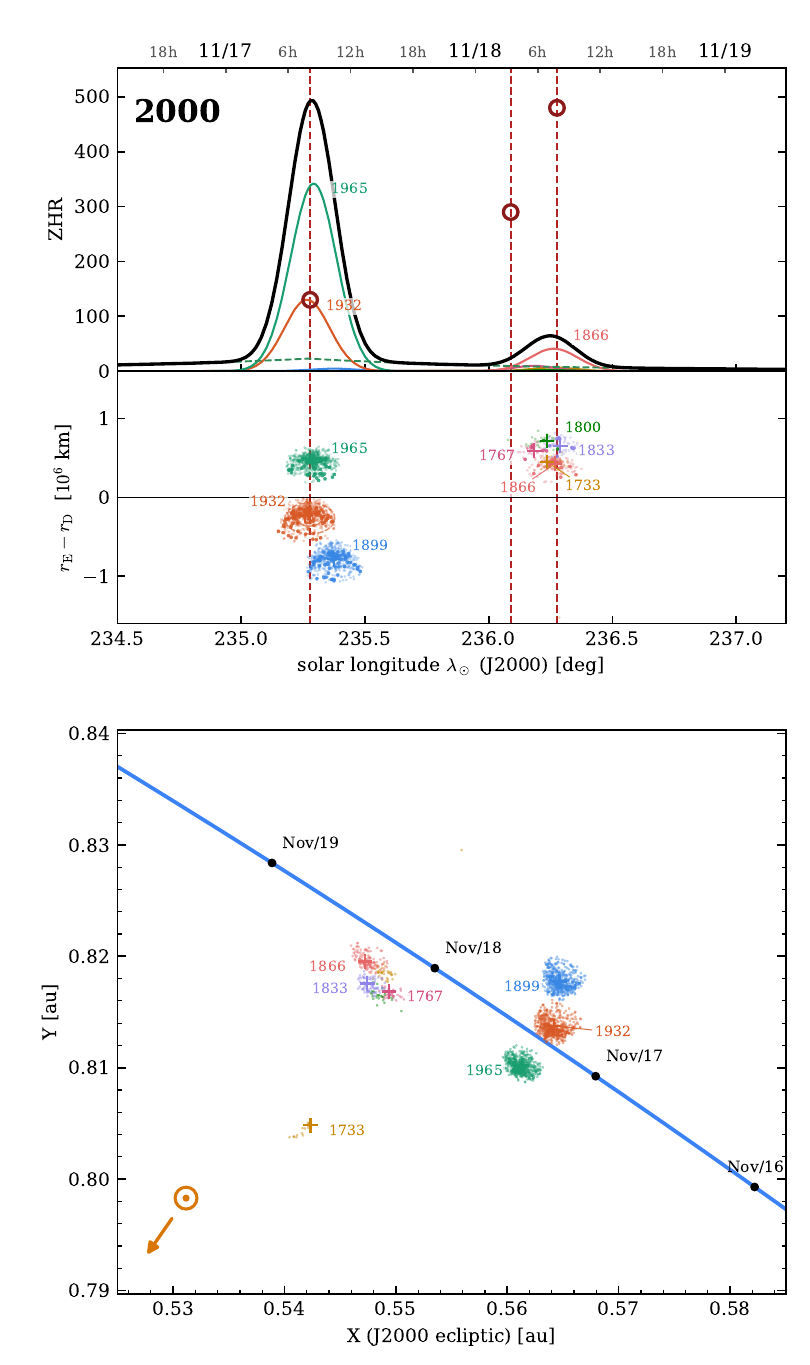}
\caption{Epoch-matched hindcast of the 2000 Leonid encounter.
The year the model fails in both directions.  It places the 1733 and
1866 sections $\simeq +3\times10^{-3}$~au inside Earth's orbit, four
times farther than \citet{asher1999imc}, and reaches only
$\times0.03$ and $\times0.09$ of the outbursts recorded at 03:24 and
07:51~UT.  The maximum it does produce from the 1932 trail on November 17 is
over-predicted eighteenfold against the ZHR $130\pm20$ measured that
night (Sect.~\ref{sec:t1932}).
Panels, symbols and conventions as in Fig.~\ref{fig:year1999}.
}
\label{fig:year2000}
\end{figure}

\begin{figure}[p]
\centering
\includegraphics[height=0.74\textheight]{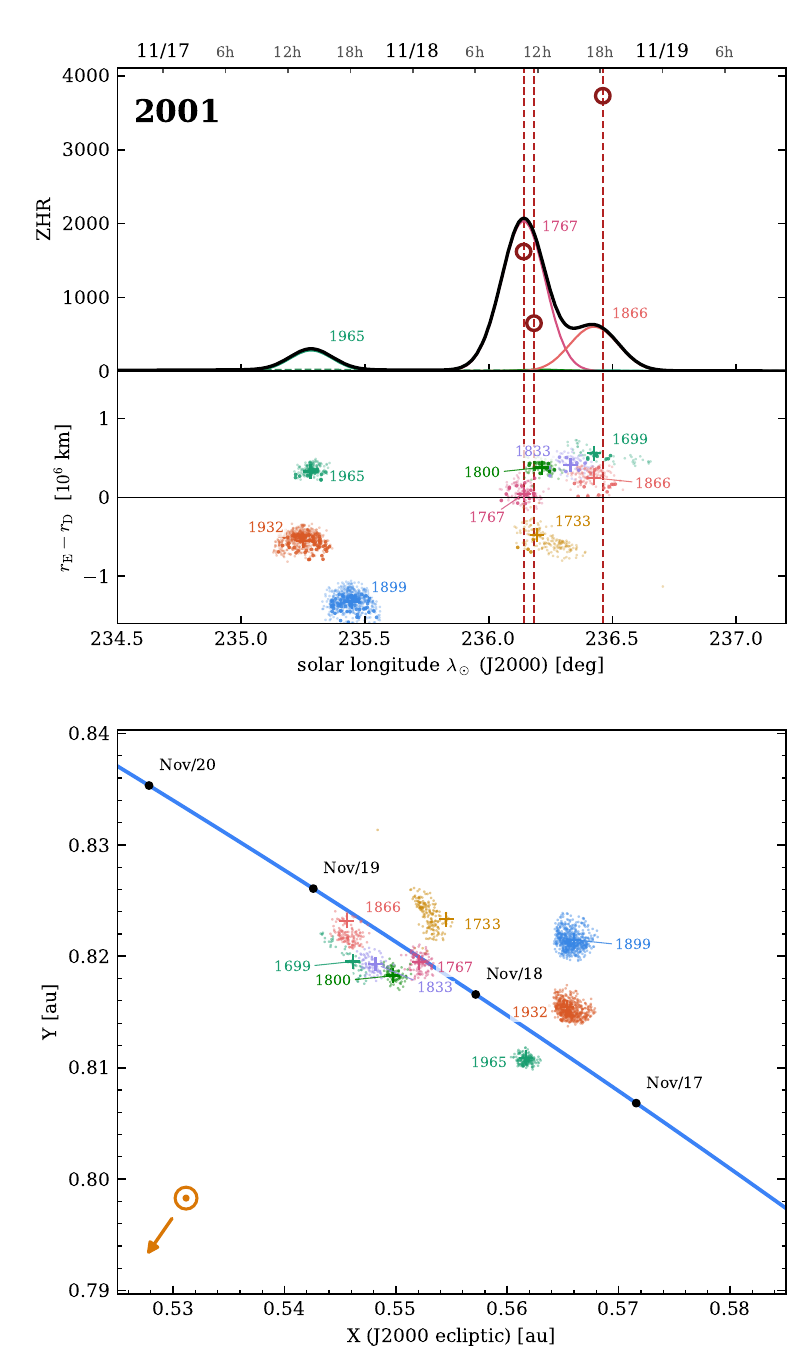}
\caption{Epoch-matched hindcast of the 2001 Leonid encounter.
The best-reproduced encounter of the sequence: the 1767 maximum
lands 1~min from the observed 10:39~UT peak at $\times1.25$ its rate.
The second observed maximum, the 1866 trail at 18:16~UT
(ZHR $3730\pm90$; \citealt{arlt2001}), is reached at only
$\times0.16$: the model places that section
$+1.7\times10^{-3}$~au inside Earth's orbit where the observed storm
requires $\lesssim10^{-3}$~au.  The six-revolution 1800 maximum that
\citeauthor{arlt2001} resolve at 11:39~UT (ZHR $650\pm40$) is missed by
a factor 31.
Panels, symbols and conventions as in Fig.~\ref{fig:year1999}.
}
\label{fig:year2001}
\end{figure}

\begin{figure}[p]
\centering
\includegraphics[height=0.74\textheight]{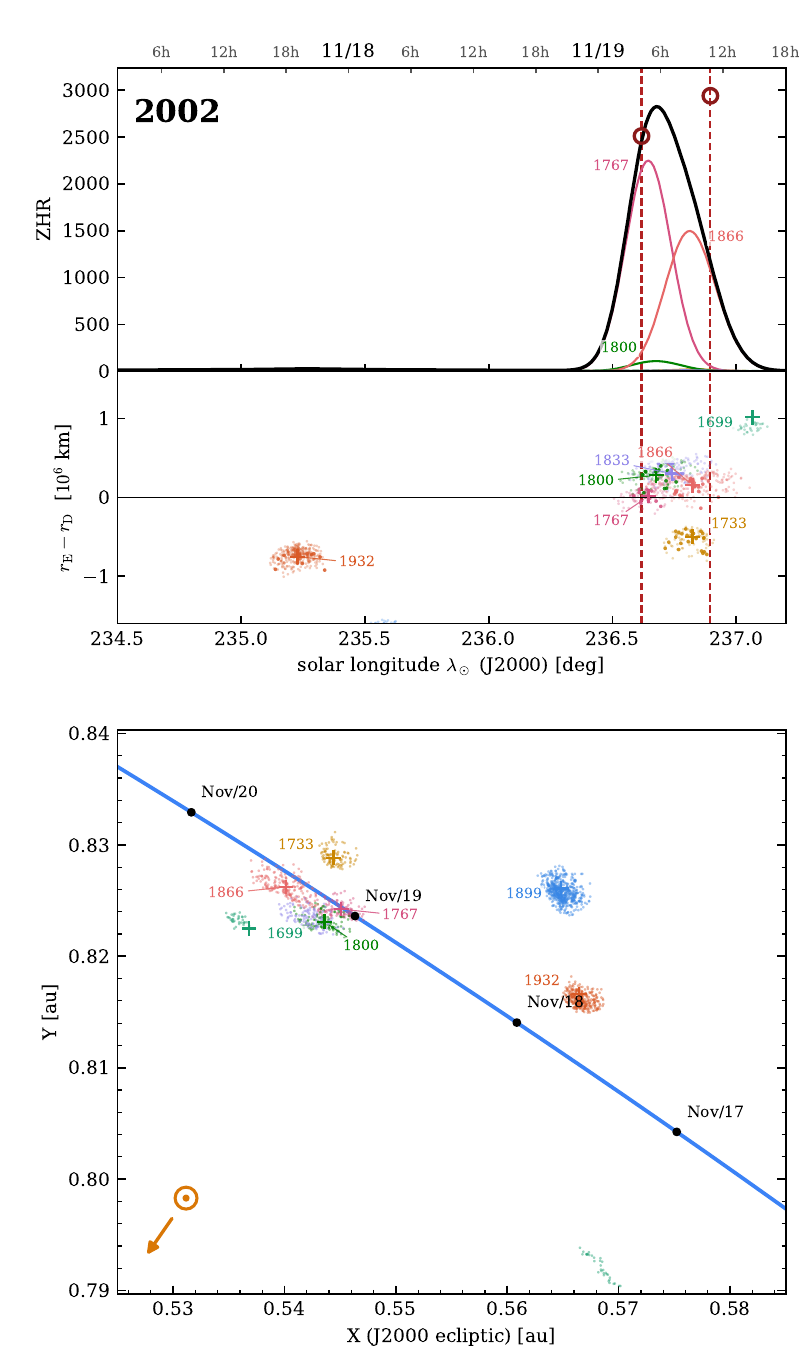}
\caption{Epoch-matched hindcast of the 2002 Leonid encounter.
The 1767 maximum is again recovered, 40~min late and at
$\times0.90$; the 1866 maximum again falls short, at $\times0.51$ from a
node $+1.0\times10^{-3}$~au inside Earth's orbit.  Observed maxima are
the IMO values of \citet{arlt2002} (ZHR $2510\pm60$ at 04:10~UT and
$2940\pm210$ at 10:47~UT); the Leonid MAC airborne campaign
\citep{abe2003} places the same pair at 04:03 and 10:49~UT.
Panels, symbols and conventions as in Fig.~\ref{fig:year1999}.
}
\label{fig:year2002}
\end{figure}

The residual uncertainty is now concentrated in the amplitude
\emph{partition} among overlapping old trails rather than in their
identity: the 1767 trail is recovered in both 2001 and 2002 with the
right timing but with amplitudes differing by a factor 1.4 between
the two years ($\times1.25$ and $\times0.90$), and the 1866 contribution
reaches only $\times0.16$ and $\times0.51$.  Since the ejection sites of these trails are now
fixed to the observed parent to $\lesssim0.15$~au, the remaining error
must accumulate in the 130--230~yr of dust propagation that follows.  We
therefore quote old-trail ($n_{\rm rev} \ge 4$) peak times with a
$\pm 3$~h systematic in addition to the class band --- halved relative to
the pre-anchoring value in line with the improvement measured in
2001/2002 --- and young-trail ($\le 3$ rev) times at $\pm 1$~h.

\subsection{The 2009 encounter: a test above the band}
\label{sec:leo2009}

Sect.~\ref{sec:limits} bounds this model to the three- to
eight-revolution band.  The 1998--2002 sequence cannot test the upper
edge of that bound, because every trail it contains lies inside it.
The 2009 return can: \citet{vaubaillon2005b} predicted enhanced
activity that November from the 1466 and 1533 trails, sixteen and
fourteen revolutions old, and \citet{koten2011} mounted a
double-station video experiment in Tajikistan to test it, with
photographic fireball data from the same return in
\citet{kokhirova2011}.  It is the only encounter in the modern record
that probes this model well above its validated range, and we ran it
for that reason: a $1.7\times10^{5}$-grain, seventeen-return
epoch-matched dataset spanning 560~yr to 1466, anchored throughout
(the pre-1600 windows through the 1366-epoch solution).

The enhancement was observed.  \citet{koten2011} place their maximum at
$22^{\rm h}07^{\rm m}\pm15^{\rm m}$~UT
($\lambda_\odot = 235\overset{\circ}{.}560$) and quote a peak flux of
$0.02\pm0.004$ meteoroids km$^{-2}$\,h$^{-1}$ to $+6.5$~mag; the IMO
visual analysis they cite gives ZHR $89\pm7$ at 20:19~UT
($\lambda_\odot = 235\overset{\circ}{.}487$), the two differing by
about two hours on a profile that was flat between 20 and 22~UT.

Our model reproduces the geometry and misses the rate, in the direction
the band predicts.  It places the 1466 encounter at 21:25~UT --- 19~min
from \citeauthor{vaubaillon2005b}'s predicted 21:44 and 42~min from
\citeauthor{koten2011}'s measured one --- but gives that section
ZHR 2.9, and a whole-profile ZHR of 20.5 at the IMO peak and 19.1 at
the video peak, of which 17.9 and 16.5 are the annual background.
Against ZHR $89\pm7$ that is $\times0.23$.  The suppression is
traceable and it is all in terms this paper has already introduced: at
$n_{\rm rev} = 16.3$ the coherence cutoff of Sect.~\ref{sec:amplitude}
contributes $f_{\rm disp} = \exp[-(16.3-12)/3] = 0.24$, the section's
node sits $+9.6\times10^{-3}$~au from Earth's orbit --- $19\sigma_r$,
far out on the Lorentzian --- and $f_{\rm M}$ is 0.074 on 75 grains.
Every one of those is an extrapolation of a term calibrated between
three and eight revolutions.

The comparison is not a vindication of this model.
\citet{vaubaillon2005b} predicted ZHR 500 where 89 was measured, an
over-prediction by $\times5.6$; we under-predict by $\times4.3$.  The two
models bracket the observation from opposite sides and neither is
close, which is what an encounter outside both models' calibrated range
should look like.  What survives on both sides is the timing, and that
is the same division --- geometry robust, amplitude not --- that
Sect.~\ref{sec:limits} draws and that \citet{egal2020review} report for
the field as a whole.  The 1533 and 1567 sections make the point more
bluntly: we place their nodes $5.9\times10^{-2}$ and
$4.1\times10^{-2}$~au from Earth's orbit and produce nothing at all,
against \citeauthor{vaubaillon2005b}'s ZHR 250 and 500, but they are
supported by 16 and 8 grains respectively and we do not regard our
placement of them as measured.

Two artefacts in the same run deserve reporting under the completeness
rule of Sect.~\ref{sec:limits}, and they fail in two different ways.

The first is the 1767 section, which produces ZHR 771 on
November 16 ($\lambda_\odot = 234\overset{\circ}{.}073$) with
$f_{\rm M} = 1.89$ just below its ceiling of 2.0 and $r = 2.15$ at the
ceiling the ejected size distribution sets --- its median grain sits
$1.22$ floors up, the signature of the wing encounter described below.  At 7.3 revolutions $f_{\rm M} \simeq
0.14$ is expected, so the estimate is $13\times$ too large.  What
distinguishes this encounter is not the trail's age but where Earth
meets it: the node lies $8.1\sigma_r$ out, in the sparse wing, where
the four-nearest-neighbour density estimator resolves a local
condensation instead of a mean linear density.  The same 1767 trail
behaves correctly wherever Earth crosses near its core --- $f_{\rm M}$
is $0.80$ and $0.95$ times the analytic $1/n_{\rm rev}$ at
$0.2\sigma_r$ in 2002 and $0.6\sigma_r$ in 2001, the two encounters it
reproduces.  Across all sections of all datasets in this paper with
$n_{\rm rev} \ge 2$ and $\ge30$ supporting grains, the ratio
$f_{\rm M}/(1/n_{\rm rev})$ has median $0.69$ and maximum $3.2$ for the
sixteen encounters within $3\sigma_r$, but scatters from $0$ to $32$ for
the ninety-nine beyond $5\sigma_r$.
The estimator is therefore trustworthy for a near-core crossing and
unconstrained in the wing, and this is a sharper criterion than the
revolution count: it is the reason the over-weighted entries of
Sect.~\ref{sec:limits} are what they are, and it is why the three
sections that carry the 2034 forecast --- all met within $2\sigma_r$
--- are not affected.

\begin{figure}[p]
\centering
\includegraphics[height=0.74\textheight]{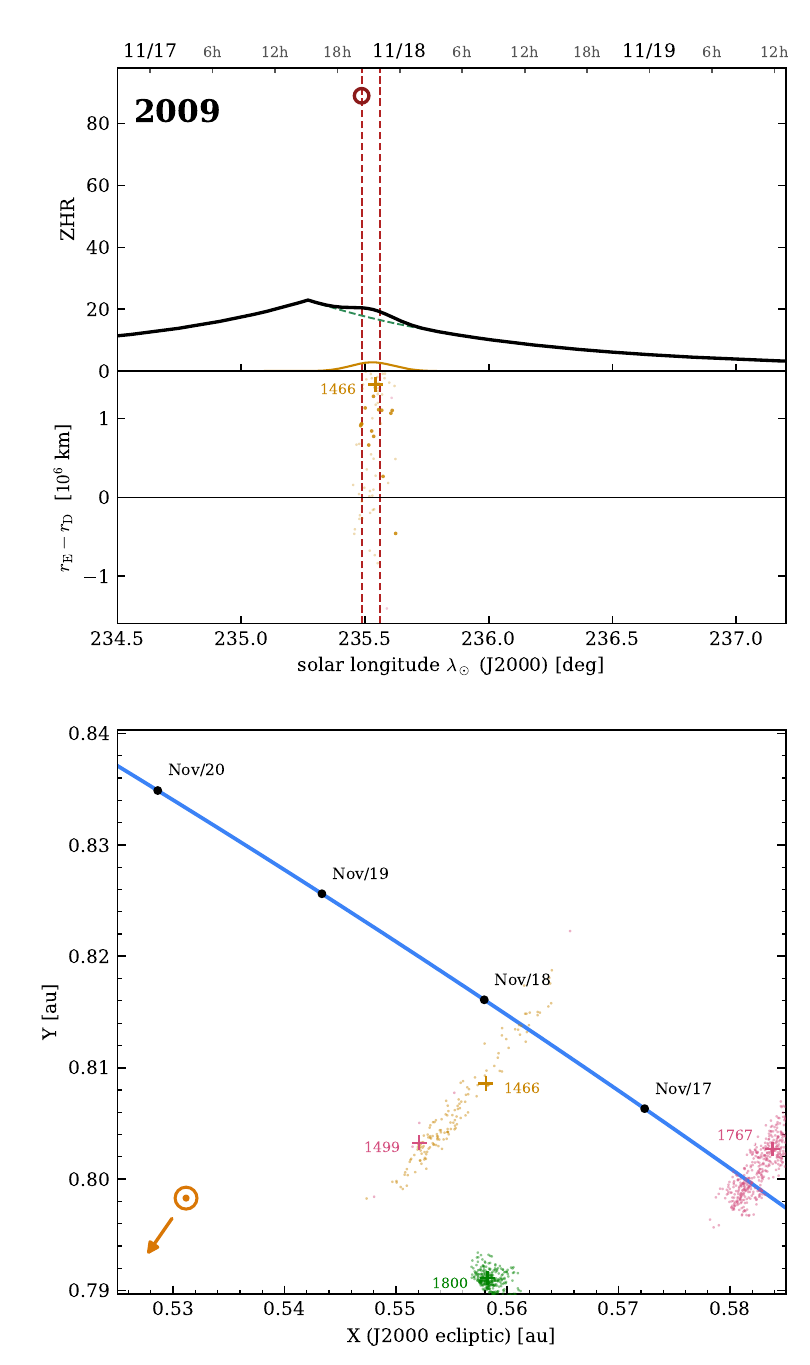}
\caption{Epoch-matched hindcast of the 2009 Leonid encounter, the only
one testing this model well above its band.  Circles and red rules mark the two published
maxima --- ZHR $89\pm7$ at 20:19~UT (IMO) and the video maximum of
\citet{koten2011} at 22:07~UT --- where the model reaches 20.5 and 19.1,
of which 17.9 and 16.5 are the annual background.
The 1466 section, sixteen revolutions old and the one
\citet{vaubaillon2005b} predicted at ZHR 500, peaks at 2.9 at 21:25~UT.
The two sections that dominate the model's output that November are
absent, and that is the point: the 1767 and 1733 artefacts peak at
$\lambda_\odot = 234\overset{\circ}{.}07$ and
$240\overset{\circ}{.}17$, outside a range held identical to the other
years.  Neither is a prediction (Sect.~\ref{sec:leo2009}).  Panels and
conventions as in Fig.~\ref{fig:year1999}.
}
\label{fig:year2009}
\end{figure}

The second is the 1733 section, and it fails for the opposite reason.
Its node is close, $+1.1\times10^{-3}$~au or $2.3\sigma_r$, but it is
supported by 13 grains.  With so few, the density estimator produces
spurious spikes wherever two or three happen to arrive together, and
the section returns several maxima across the window rather than one.
This is also the encounter that exposed an error in how the profile
was assembled.  The temporal shape was originally built from the
along-trail density alone, so a trail crossing the ecliptic at more
than one epoch acquired one bump per crossing, and every bump inherited
the amplitude computed at the \emph{closest} crossing --- the model
reported a rate at times when Earth was nowhere near the trail.  The
1733 grains all lie at the November 22 crossing, yet a secondary
maximum of ZHR $1.0\times10^{2}$ appeared on November 19, where the
ecliptic-plane panel of Fig.~\ref{fig:year2009} shows no grains at all.
Two things were wrong.  The shape ignored the radial miss distance, so
every bump inherited the amplitude of the closest crossing; weighting it
by the same Lorentzian the cross-section uses removes that.  And the
along-trail bandwidth carried an edge floor $h \ge |\Delta t|/3$, which
holds the Gaussian exponent fixed at $\mathrm{e}^{-4.5}$ for every
$|\Delta t| \ge 3\sigma_t$ and leaves the density decaying only as
$\sigma_t/h \sim 1/|\Delta t|$ --- a power law, so a grain whose nodal
passage is months from Earth's arrival still contributed a per cent of
an on-time one, and on a sparse trail each such grain raised its own
bump.  Capping the floor at $\sigma_t$ restores Gaussian decay.  With
both corrections the 1733 section returns a single maximum instead of
eight, its November 19 residual falls to ZHR 0.03 against an annual
background of 5.3 there, and three spurious entries elsewhere --- a second 1433
maximum in 2030 and second 1733 maxima in 2033 and 2035 --- disappear.
The corrections change the overall scale, so $Z_{\rm storm}$ was
re-derived on the same three storms, giving $8.6\times10^{3}$ in place
of $8.3\times10^{3}$; the 2034 forecast is invariant to the cap once
that is done, moving by less than 1\% across caps from $\sigma_t$ to
$3\sigma_t$.  All results in this paper use the corrected form.
What the correction cannot repair is the support: thirteen grains is
below the count at which we quote any rate, so the section's surviving
November 22 maximum should be read as noise rather than as a prediction.

Neither artefact has an observational counterpart, and neither is
predicted by any published model: \citet{vaubaillon2005b} place all
2009 activity on November 17, as does \citet{maslov2007}, and the
record for that return holds nothing above ZHR $89\pm7$ on any night.
A ZHR 771 event 29~h before the annual maximum would have
been the meteor event of the decade.  We report them because a table of
nominated trails would not have shown them.

\clearpage

\section{Leonid forecasts for 2031--2035}
\label{sec:forecast30}
\label{sec:leonids}

\begin{figure}[p]
\centering
\includegraphics[height=0.74\textheight]{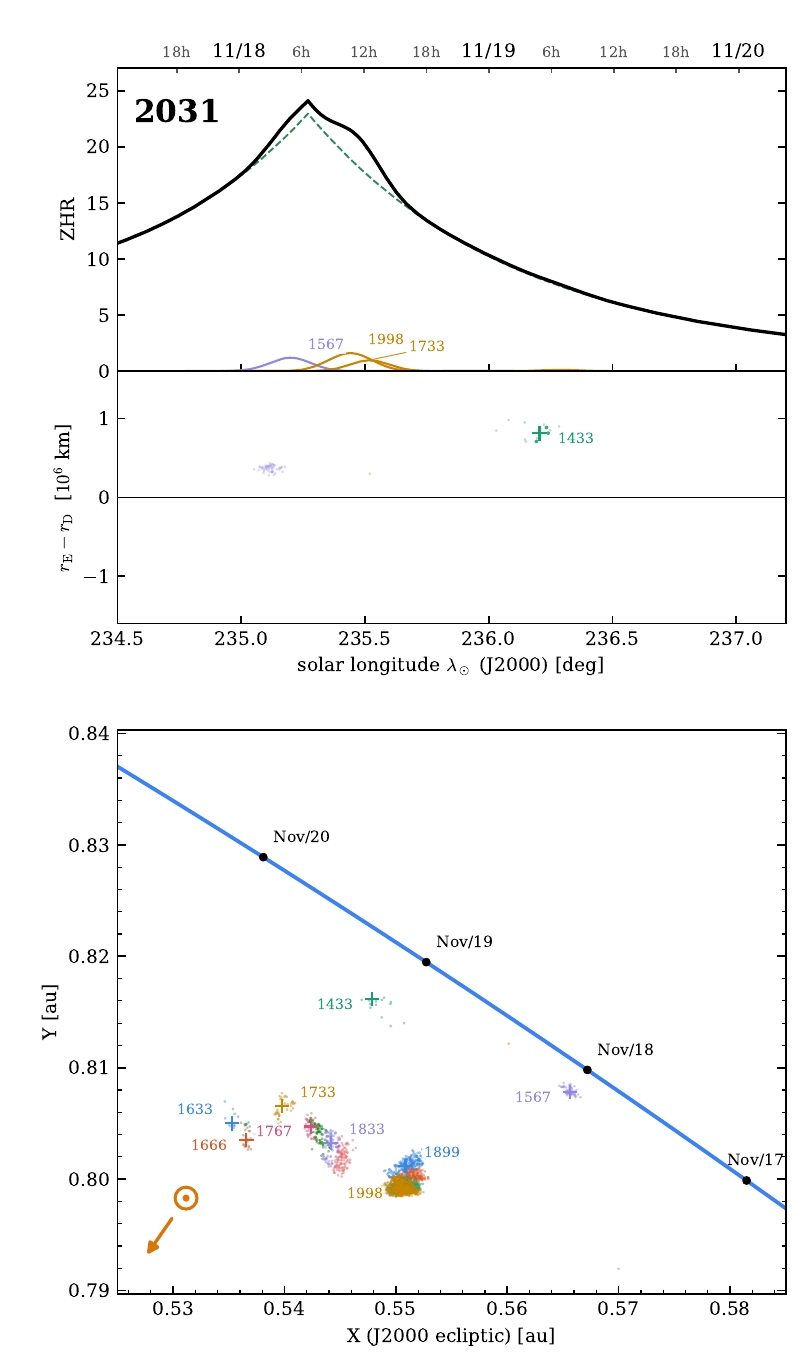}
\caption{Leonid forecast for 2031.  Panels as in Fig.~\ref{fig:year1999}; the class-B band
spans a factor 12 and is omitted here, being quoted in
Table~\ref{tab:leonids}.
A quiet year: no coherent trail section reaches Earth's orbit, and the profile is the annual background alone, ZHR 24.}
\label{fig:year2031}
\end{figure}

\begin{figure}[p]
\centering
\includegraphics[height=0.74\textheight]{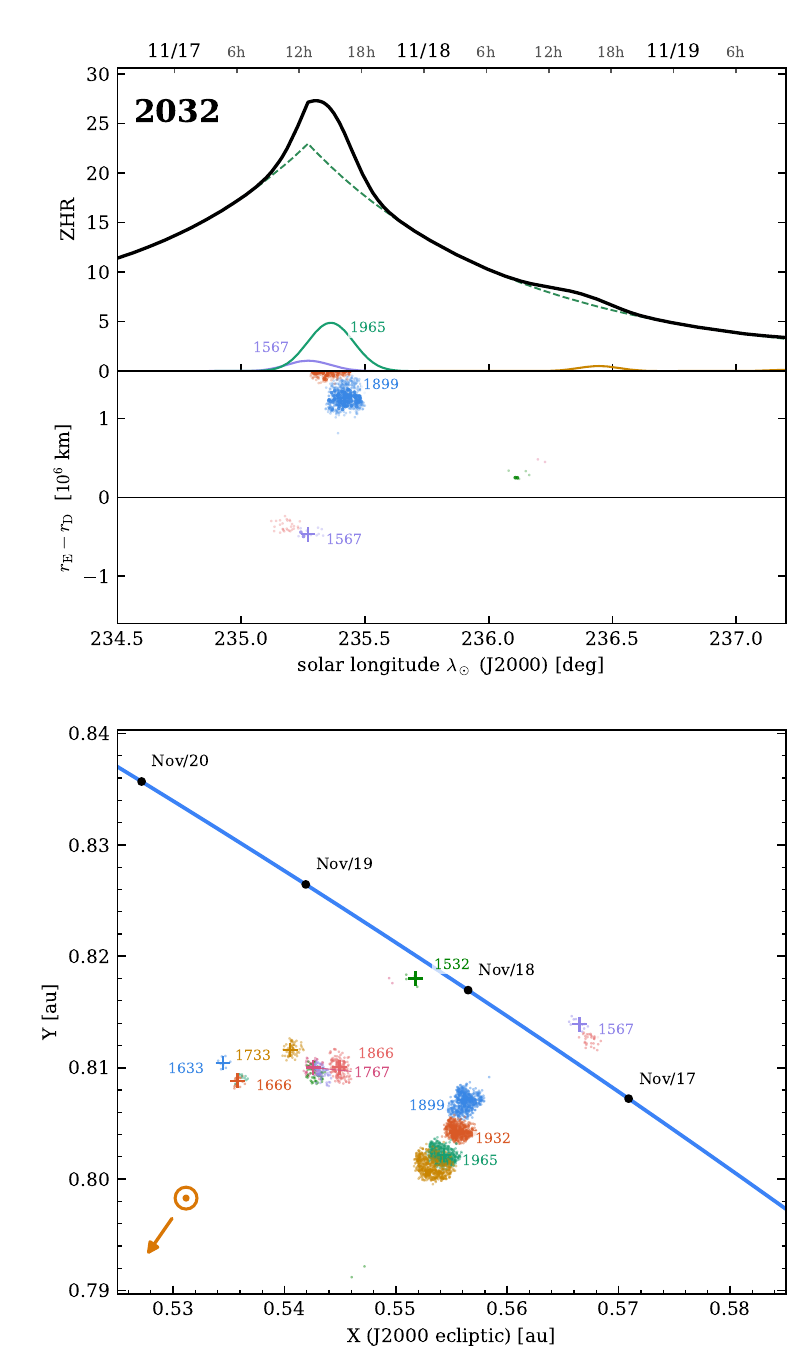}
\caption{Leonid forecast for 2032.  Panels and conventions as in Fig.~\ref{fig:year2031}.
Nothing reaches Earth: the closest section is the twelve-revolution 1567 trail at $3.1\times10^{-3}$~au ($6.2\sigma_r$) on ten grains, and the profile stays at the annual background, ZHR 27.  The frozen-element version of this forecast put the 1899 trail $2.7\times10^{-3}$~au inside Earth's orbit and lifted the rate to 62; at its own epoch that section is $1.4\times10^{-2}$~au away.}
\label{fig:year2032}
\end{figure}

\begin{figure}[p]
\centering
\includegraphics[height=0.74\textheight]{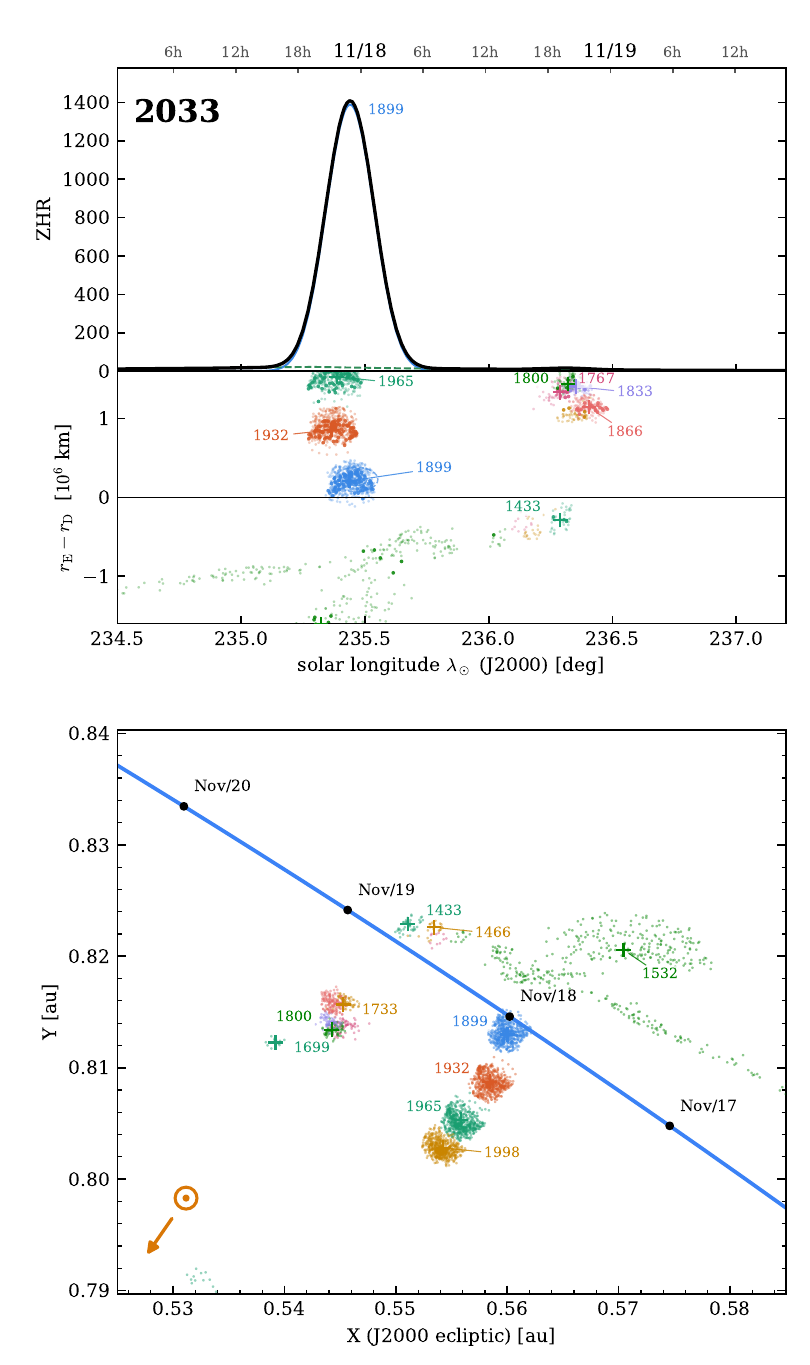}
\caption{Leonid forecast for 2033.  Panels and conventions as in Fig.~\ref{fig:year2031}.
The strongest encounter of the decade after 2034, and one the frozen-element forecast missed almost entirely.  The four-revolution 1899 trail --- the same trail that produced the 1999 storm --- passes $1.4\times10^{-3}$~au inside Earth's orbit ($2.8\sigma_r$) on 607 grains and gives ZHR $1.4\times10^{3}$ at 22:56~UT on November 17.  The earlier version of this forecast placed the same section further out and quoted ZHR 130.}
\label{fig:year2033}
\end{figure}

\begin{figure}[p]
\centering
\includegraphics[height=0.74\textheight]{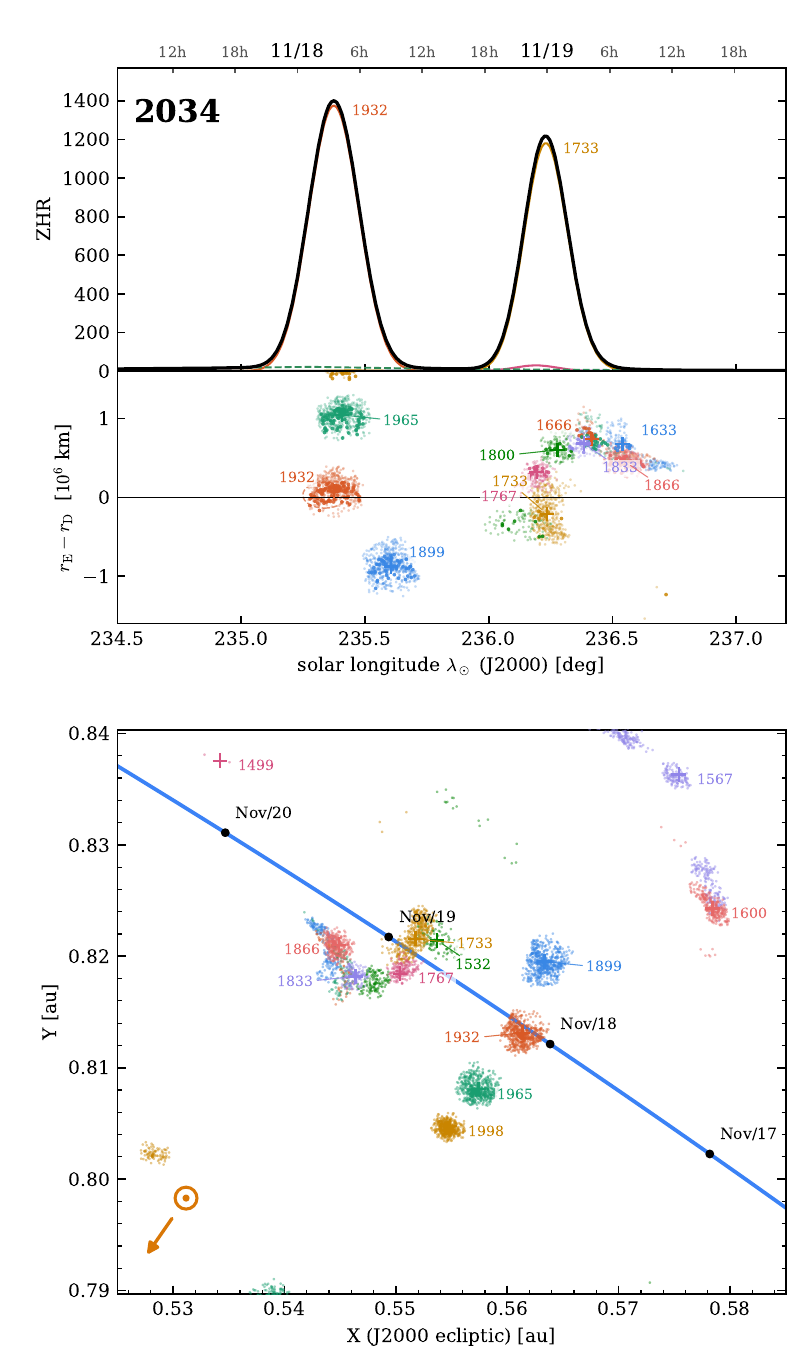}
\caption{Leonid forecast for 2034.  Panels and conventions as in Fig.~\ref{fig:year2031}.
The decade's key encounter, and two maxima of comparable height twenty hours apart.  The three-revolution 1932 cloud rides the Earth line early on November 18 ($+0.5\times10^{-3}$~au, $1.1\sigma_r$, 913 grains) and gives ZHR $1.4\times10^{3}$ at 03:30~UT with the empirical density factor of Sect.~\ref{sec:t1932} applied; the nine-revolution 1733 section crosses at $-1.4\times10^{-3}$~au late the same night and gives ZHR $1.2\times10^{3}$ at 23:45~UT.  The eight-revolution 1767 trail, which the frozen-element version of this forecast made the storm of the night at ZHR $3.5\times10^{3}$, sits $4.3\sigma_r$ out here and contributes 30.}
\label{fig:year2034}
\end{figure}

\begin{figure}[p]
\centering
\includegraphics[height=0.74\textheight]{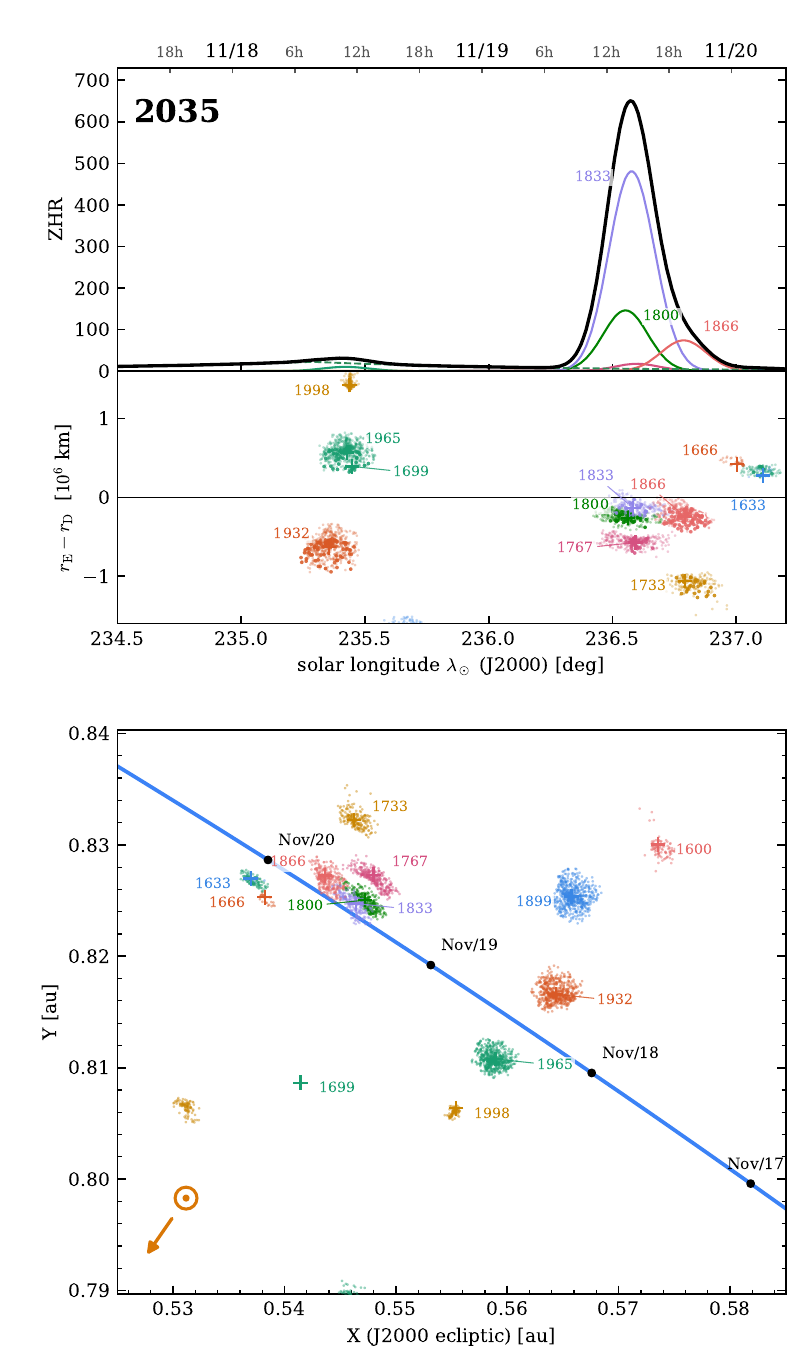}
\caption{Leonid forecast for 2035.  Panels and conventions as in Fig.~\ref{fig:year2031}.
A third active year.  The six-revolution 1833 trail passes $0.9\times10^{-3}$~au inside Earth's orbit ($1.8\sigma_r$) and gives ZHR 480 at 14:16~UT on November 19, with the 1800 and 1866 sections adding 146 and 74 within six hours; the summed maximum is ZHR 651.}
\label{fig:year2035}
\end{figure}

The forecast rests on five datasets, one per year, each carrying
450\,000 grains in eighteen trails (1433--1998; 25\,000 grains per
trail) at that year's own encounter epoch.  They come from a single
N-body integration per return, written out as it passes each November
(Sect.~\ref{sec:systematics}); the alternative --- one dataset at the
present epoch, read at every later year with the elements frozen --- omits
$13\sigma_r$ of node motion and is what an earlier version of this
forecast did.  Generation follows Sect.~\ref{sec:stream} with all
ejection sites anchored to the JPL ephemeris --- the six pre-1600
returns through the 1366-epoch orbit solution, which Horizons
propagates before the modern solution's 1599 floor --- and the
calibration is as above
($Z_{\rm storm} = 8.6\times10^{3}$, $\sigma_r = 5\times10^{-4}$~au, node
correction $-0.11^\circ$ for $n_{\rm rev}\ge4$).  The forecast begins at
the perihelion return: 55P reaches perihelion on 2031 May 20, and the
2030 encounter precedes it.  The integrations do write a 2030 snapshot,
which Sect.~\ref{sec:systematics} uses to measure the frozen-element
systematic, but that year is not forecast here --- it produces the annual
background and nothing else.
Table~\ref{tab:leonids} lists the annual forecasts, and
Figs.~\ref{fig:year2031}--\ref{fig:year2035} show each year in turn:
its ZHR profile, its nodal cross-section and its ecliptic-plane
geometry.

\begin{table}[t]
\centering
\small
\caption{Leonid forecasts, 2031--2035, from the ephemeris-anchored
$4.5\times10^{5}$-grain dataset (confidence class B: 0.29--3.5).  Times
are UT at the model peak; old-trail ($n_{\rm rev}\ge4$) times carry an
additional $\pm3$~h systematic (Sect.~\ref{sec:systematics}).  ZHR is the
total profile (trails plus annual background); the band is the class-B
interval.  Up to the three strongest maxima are listed per year; the 2034 maxima are resolved on the 15-min grid.}
\label{tab:leonids}
\begin{tabular}{lrlll}
\toprule
Year & Peak ZHR & Band & Peak time & Leading trails (ZHR) \\
\midrule
2031 & 24 & 23--34 & Nov 18 06:36 & background only \\
2032 & 27 & 23--74 & Nov 17 13:35 & 1965 (4), 1567 (1) \\
2033 & 1409 & 158--13917 & Nov 17 22:56 & 1899 (1389) \\
2034a & 1401 & 158--13824 & Nov 18 03:30 & 1932 (1376), 1998 (4), 1965 (1) \\
2034b & 1217 & 129--12099 & Nov 18 23:45 & 1733 (1180), 1767 (28) \\
2035 & 651 & 70--6457 & Nov 19 14:16 & 1833 (480), 1800 (143), 1767 (17) \\
\bottomrule
\end{tabular}
\end{table}

\paragraph{2034: two maxima, twenty hours apart}
The decade's strongest activity is predicted for 2034 November 18, and it
is a night with two comparable maxima rather than one storm.  The first
is at 03:30~UT, from the three-revolution 1932 trail
($r_{\rm E}-r_{\rm D} = +5.4\times10^{-4}$~au, $1.1\sigma_r$, 913
supporting grains): a near-direct hit giving ZHR $1.4\times10^{3}$ once
the empirical density factor of Sect.~\ref{sec:t1932} is applied, and
ZHR $2.6\times10^{4}$ without it.  Its time carries the $\pm1$~h
young-trail systematic.

Twenty hours later, at 23:45~UT, Earth crosses the nine-revolution 1733
trail ($r_{\rm E}-r_{\rm D} = -1.4\times10^{-3}$~au, $2.8\sigma_r$, 283
grains) for ZHR $1.2\times10^{3}$, with the $\pm3$~h old-trail
systematic on its time.  The summed profile reaches ZHR 1401 and 1217 at
the two maxima (Table~\ref{tab:leonids}).  \emph{Which of the two
dominates is the sharpest observational test this model offers}: with
the density correction they are comparable, without it the early 1932
maximum outweighs the later one twentyfold.  The test turns on a single
trail's density and not on any geometric quantity, since neither time
nor node distance depends on the correction.

The eight-revolution 1767 trail, which an earlier version of this
forecast made the storm of the night at ZHR $3.5\times10^{3}$ and 22:30~UT,
is not part of the picture: computed at its own epoch it lies
$2.1\times10^{-3}$~au inside Earth's orbit ($4.3\sigma_r$) and reaches
ZHR 30 at 23:00~UT.  That reversal is the frozen-element systematic of
Sect.~\ref{sec:systematics} in its most consequential form --- the
extrapolated dataset moved every node inward by several $\sigma_r$, and
the 1767 section happened to land on Earth's orbit as a result.  The
five-revolution 1866 trail, the third maximum of that earlier forecast,
likewise falls to ZHR 3.2 at $6\sigma_r$.  The waxing moon (first
quarter Nov 17) sets before the radiant culminates for Atlantic
longitudes at the 03:30 encounter.
\paragraph{Sections whose population index cannot be measured}
Some sections carrying a forecast are floor-bound in the
sense of Sect.~\ref{sec:maglim}: their grains sit against the sampler's
smallest size, so the slope there is set by the truncation and the model
takes the ceiling rather than a measurement.  For those the rate is
quoted as a range.  The upper end is the model as specified; the lower
end replaces the ceiling with the median index of the sections the same
dataset \emph{does} measure, leaving everything else --- node, $f_{\rm M}$,
coherence, calibration --- untouched.  One section of this forecast is
affected: the 1932 section of 2034, whose rate is 436--1376.

Nothing else moves by more than 1\%.  The 1899 section of 2033, the 1833
and 1800 sections of 2035 and the 1733 section of 2034 all have
measured indices, and the same is true of the three storms that fix
$Z_{\rm storm}$, so the amplitude scale is unaffected.  For 2034 the
width of the range matters less than what it does to the ordering: at the
lower end the 1932 maximum falls below the 1733 one, so which of that
night's two maxima dominates depends on the index treatment as well as on
the density factor of Sect.~\ref{sec:t1932}.  We report both rather than
choose, because the observation will separate them.

\paragraph{Comparison with the published forecast}
Two trail-by-trail Leonid forecasts covering this return have been
published --- \citet{vaubaillon2005b}, whose Table 5 tabulates Leonid
encounters from 2006 to 2100, and \citet{maslov2007} --- and
Table~\ref{tab:cmp} sets both beside ours.  The comparison is more
informative than a single number would be, because the three models
share nothing but the target: different parent integrations, ejection
laws, encounter kernels and calibration data.  \citet{vaubaillon2005b}
in particular weight each simulated particle by the parent's dust
production and quote an absolute flux, so their rates are derived
independently of any storm calibration.

They agree on one encounter and disagree on everything else.  The
three-revolution 1932 trail is nominated by all three models and they
place it within 30~min --- 03:04, 03:30 and 03:34~UT --- with rates
spanning a factor 5 (400--500, 1376, 2000).  For a section computed a
century after ejection from three independent parent integrations, that
is a genuine mutual confirmation, and it is consistent with the review
finding that peak times are the robust output of trail models while
rates are not \citep{egal2020review}.

Beyond that trail the three forecasts describe different nights.  Our
second maximum, the nine-revolution 1733 section at 23:45~UT and ZHR
$1.2\times10^{3}$, appears in neither published forecast.  Conversely
three of the four sections that carry \citeauthor{vaubaillon2005b}'s
November 19 --- 1633, 1666 and 1866, at ZHR 350, 1350 and 20 --- lie
$9.0$, $9.9$ and $6.0\sigma_r$ from Earth's orbit in our epoch-matched
dataset and contribute 0.35, 0.15 and 3.2 between them, and the
eight-revolution 1767 trail he puts at ZHR $1.05\times10^{3}$ we put
$4.3\sigma_r$ out at ZHR 30.  The fourth, the ten-revolution 1699 trail
he puts at ZHR 1500, we report as undetermined rather than absent: it is
the one section whose answer flips with the ejection realisation, from
$63\sigma_r$ and nothing to $3.8\sigma_r$ and ZHR 132 (note~$d$ of
Table~\ref{tab:cmp}).
\citeauthor{maslov2007} expects ZHR 30--40 from the 1899 trail at
09:02~UT where we place that section $5.9\times10^{-3}$~au inside
Earth's orbit.

We do not present our values as the safe ones.  Two independently built
models agreeing against us on the November 19 sections, which we place
an order of magnitude further out, is a warning rather than a
vindication --- and our own November 18 maxima rest on 913 and 283
grains in trails whose realisation-to-realisation spread
(Sect.~\ref{sec:calib}) is a factor 1.5--2.  What can be said is that
the disagreements are geometric and dated, so a single night settles
them: an early maximum near 03:30~UT is common ground, activity late on
November 18 is ours alone, and activity on November 19 is theirs alone.

One further agreement deserves emphasis because it tests a part of the
model nothing else does.  \citeauthor{maslov2007} annotates each
encounter with an expected brightness, derived from his own treatment
of the ejection offset: the 1932 encounter is ``significantly lower
than average'' in brightness with radio rates stronger, while the 1767
encounter is ``close to the average level''.  Our per-section
population index, measured independently from the local grain-size
distribution (Sect.~\ref{sec:maglim}), gives $r = 2.38$ for the 1932
section against $r = 1.38$ for the 1767 section, on a scale where the
three calibrating storm sections lie at $1.33$--$1.47$.  Two
unrelated methods thus identify the same trail as the faint-rich one
and the same trail as the normal one.  This is the only external
corroboration the per-section conversion of Sect.~\ref{sec:maglim}
has, and it also supports the physical reading of the 1932 excess in
Sect.~\ref{sec:t1932}: the trail really is depleted in visually
detectable meteoroids, which is what the model cannot represent and
what the empirical density factor absorbs.

\paragraph{The quiet years}
2031 and 2032 are the annual background and nothing else: no trail of the
eighteen comes within $6\sigma_r$ of Earth's orbit in 2031, and the
strongest section of 2032 is the two-revolution 1965 trail at
$27\sigma_r$, contributing ZHR 5 to a total of 27.

Two encounters of the pre-anchoring version of this model do not
survive, and both are instructive.  The 1866/1767 complex that made 2033
a storm there (predicted ZHR $1.2\times10^{4}$) passes $7.7$ and
$8.9\times10^{-3}$~au from Earth's orbit once the ejection sites are
anchored and the year is computed at its own epoch --- 15 and
18$\sigma_r$ --- and contributes ZHR 0.5 and 0.4.  What makes 2033 an
active year instead is the four-revolution 1899 trail, the same trail
that produced the 1999 storm, at $+1.4\times10^{-3}$~au ($2.8\sigma_r$,
607 grains) for ZHR 1389 at 22:56~UT.  The earlier storm prediction was
therefore an artefact of the 55P back-integration error documented in
Fig.~\ref{fig:anchor}, a cautionary result since that error is invisible
to any internal convergence or resolution test.  The second is the
eighteen-revolution 1433 trail, newly included in this dataset: an
earlier, frozen-element version of this forecast placed it within
$1\times10^{-3}$~au of Earth's orbit and gave it a maximum of its own,
whereas computed at its own epoch its closest approach in these five
years is $1.9\times10^{-3}$~au ($3.9\sigma_r$, in 2033) and its largest
contribution anywhere is the ZHR 7 that goes with it.  Old-trail
encounters are nonetheless worth reporting: \citet{vaubaillon2005b} predicted 2009
November 17 activity from the 1466 and 1533 trails --- the latter is the
section our dataset carries as 1532 --- and \citet{koten2011} confirmed
it with a double-station video campaign.

\clearpage
\begin{table}[t]
\centering
\footnotesize
\caption{The 2034 November encounter as predicted by the two published
trail-by-trail forecasts covering this return and by the present model,
now computed from the epoch-matched $4.5\times10^{5}$-grain dataset of
Sect.~\ref{sec:stream}.  Every row that any of the three places above
ZHR 20 is listed, in the completeness spirit of
Table~\ref{tab:leo9802}, in time order.  The horizontal rule separates
November 18 from November 19.  The three models share nothing ---
different parent integrations, ejection laws, encounter kernels and
calibration data; \citet{vaubaillon2005b} in particular weight simulated
particles by the parent's dust production and quote an absolute flux,
whereas ours buys its amplitude from one measured storm.  The
three-revolution 1932 trail is the one encounter all three nominate and
it is the one they agree on: the maxima span 30~min and the rates a
factor 5.  Everything else disagrees.  Our nine-revolution 1733 section
--- the second maximum of the night at ZHR $1.2\times10^{3}$ --- appears
in neither published forecast, while the 1633, 1666 and 1699 sections
that carry \citeauthor{vaubaillon2005b}'s November 19 activity sit
$9$--$63\sigma_r$ from Earth's orbit in ours and contribute nothing, and
the eight-revolution 1767 trail he places at ZHR $1.05\times10^{3}$ we
place $4.3\sigma_r$ out at ZHR 30.  These are clean, dated, falsifiable
differences, and the night discriminates between the three models rather
than testing ours alone.  The last column is our nodal distance in the
convention of Table~\ref{tab:leo9802} (positive = trail inside Earth's
orbit) with its width in units of $\sigma_r = 5\times10^{-4}$~au.}
\label{tab:cmp}
\begin{tabular}{l r ll ll ll r}
\toprule
Trail & $n_{\rm rev}$ &
\multicolumn{2}{c}{\citet{vaubaillon2005b}} &
\multicolumn{2}{c}{\citet{maslov2007}} &
\multicolumn{2}{c}{This model} & $r_{\rm E}-r_{\rm D}$ \\
\cmidrule(lr){3-4}\cmidrule(lr){5-6}\cmidrule(lr){7-8}
 & & UT & ZHR & UT & ZHR & UT & ZHR & [$10^{-3}$~au] \\
\midrule
1932 & 3.1 & Nov 18 03:34 & 2000 & Nov 18 03:04 & 400--500 & Nov 18 03:30 & 1376 & $+0.54$ \\
1899 & 4.1 & --- & --- & Nov 18 09:02 & 30--40 & --- & --- & $-5.86$ \\
1433 & 18.1 & Nov 18 13:13 & 45 & --- & --- & --- & --- & $-14.04$ \\
1767 & 8.0 & Nov 18 22:43 & 1050 & Nov 18 22:04 & 150--250 & Nov 18 23:00 & 30 & $+2.14$ \\
1733 & 9.0 & --- & --- & --- & --- & Nov 18 23:45 & 1180 & $-1.40$ \\
\midrule
1800 & 7.0 & Nov 19 00:43 & 70 & --- & --- & --- & --- & $+4.02$ \\
1833 & 6.0 & Nov 19 02:37 & 90 & --- & --- & --- & --- & $+4.56$ \\
1633 & 12.1 & Nov 19 03:37 & 350 & --- & --- & --- & --- & $+4.52$ \\
1666 & 11.1 & Nov 19 04:39 & 1350 & --- & --- & --- & --- & $+4.96$ \\
1699 & 10.1 & Nov 19 05:46 & 1500 & \multicolumn{2}{c}{(blended with 1866)} & ---$^{d}$ & ---$^{d}$ & $+31.54$ \\
1866 & 5.1 & Nov 19 05:45 & 20 & Nov 19 05--06 & 300--400$^{a}$ & --- & --- & $+3.00$ \\
\bottomrule
\end{tabular}
\\[2pt]
\begin{minipage}{0.92\linewidth}\footnotesize
$^{a}$ \citet{maslov2007} treats the 1699 and 1866 sections as partially
superimposed and quotes one rate for the pair.
$^{b}$ a dash in our columns means the section produces less than ZHR 5:
1866 reaches 3.2, 1833 2.5, 1800 0.8, 1633 0.35, 1666 0.15 and 1433
nothing at all.  Every one of them is a miss by $6\sigma_r$ or more, not a
rate the calibration could plausibly rescue.  The 1699 entry is a
different case; see note $d$.
$^{c}$ every entry is a single trail, matching what the published
forecasts tabulate; Table~\ref{tab:leonids} instead lists the summed
maxima an observer would see, ZHR 1401 and 1217.
$^{d}$ the one entry we report as undetermined rather than absent: with a
different ejection realisation this section sits $1.9\times10^{-3}$~au from
Earth's orbit ($3.8\sigma_r$) and gives ZHR 132, where the dataset adopted
here puts it $63\sigma_r$ out and gives nothing.  The cause is the
discretisation of the ejection into at most 40 bursts per return, not the
frozen-element systematic (Sect.~\ref{sec:calib}), so we make no claim
against \citeauthor{vaubaillon2005b}'s ZHR 1500 here.
\end{minipage}
\end{table}
\clearpage

\section{Discussion}
\label{sec:discussion}
The central methodological result of this work is negative in form and
practical in consequence: for a trail model of this class, the parent's
historical trajectory --- not the ejection law, the size distribution or
the encounter kernel --- sets the accuracy floor, and it does so in a way
that internal consistency checks cannot detect.  The pre-anchoring
version of this model passed a $4\times10^4$-grain convergence study, an
8-dimensional orbit-covariance Monte Carlo, and a resolution study of
the encountered trail section, yet predicted a $10^4$ storm for 2033
that disappears once the ejection sites are placed where the observed
comet actually was.  Convergence in particle number certifies the
sampling, not the geometry.  The remedy we adopt is cheap and, we would
argue, should be standard: the ejection geometry is a solved problem for
any comet with a Horizons ephemeris, and there is no reason for a trail
model to re-derive it from a single modern orbit solution.
A second, independent lesson concerns how such a model is scored.  The
customary practice --- comparing the model against the trails the
classical analyses nominate --- measures only the reproduction of known
events, and we found it actively misleading.  Scored that way our
1998--2002 sequence returns ratios between $\times0.89$ and
$\times1.25$; scored on \emph{every} maximum the model generates
(Sect.~\ref{sec:limits}), the same runs are seen to place
ZHR $9.5\times10^{3}$ in 1999 and ZHR $2.4\times10^{3}$ in 2000 on the
two-revolution 1932 trail, where the record holds a sub-feature and a
ZHR $130\pm20$ outburst respectively, and the summed 1999 profile
over-predicts by a factor 3.3.  For a model whose purpose is to forecast
rather than to explain, the false-positive rate is as much a figure of
merit as the hit rate, and it costs one extra table to report.

Dissecting those excesses (Sect.~\ref{sec:t1932}) yields the third
result, and the one with the widest reach.  Every over-predicted
encounter in the record is a \emph{near miss} --- a section two to three
$\sigma_r$ from Earth's orbit --- and every reproduced storm is a
near-direct hit.  On
the steep flank of the radial profile a node error of
$5\times10^{-4}$~au, which is negligible for a direct hit and comparable
to the residual we already document for old trails, changes the predicted
rate by a factor 2.2.  The practical consequence is that the uncertainty
of a trail prediction is not a single number per model: it depends on
where the encounter falls on the cross-section, and a near-miss forecast
deserves a wider interval than a direct hit computed by the same code
from the same data.  Our class bands are per shower, which is the coarsest
useful granularity; making them per encounter would be a natural
refinement.
The limits that remain are those anchoring cannot reach.  (i) The dust,
once released, is propagated by our own integrator for up to three
centuries; the surviving $-0.11^\circ$ node offset and the missing
1866-trail component of 2001/2002 must originate there, and would
require either a higher-fidelity dust integration or --- as
\citet{mcnaught1999} did empirically --- a fitted per-trail offset.
(i\emph{a}) The absolute density factor $f_{\rm M}$ is reliable in the
band the calibration storms occupy --- it matches \citet{lyytinen2000} to
3--13\% at seven revolutions --- but is $1.7$ times too large at two
revolutions, where a linear-stretching normalisation is applied to a
trail that has not yet stretched into a smooth tube.  (i\emph{b}) The
model carries one ejection size distribution per trail and therefore
cannot reproduce the mass sorting along a trail: the section of the 1932
trail met in 2000 is, in \citeauthor{lyytinen2000}'s words, ``mostly
faint'', and no single-population model will get its \emph{visual} rate
right.
(ii) Horizons integrates comets no earlier than 1599, so trails older
than four centuries fall back to the internal back-integration, with the
error budget of Fig.~\ref{fig:anchor}; for the lost comet
289P/Blanpain, where no modern solution can be trusted backward at all,
we find that only forward integration from the historical 1819 orbit
places trails near Earth.  (iii) The frozen-element forecast requires
epoch-matched datasets for precision hindcasting: run from a dataset
whose epoch is several revolutions away from the encounter, the frozen
elements drift, the staleness guard of Sect.~\ref{sec:conf} demotes the
result to class~C, and the rate can move by more than an order of
magnitude.  (iv) The amplitude calibration
rests on a handful of storms per shower; assimilating continuous
optical-flux series \citep{vida2022} would replace the anchor with a
fitted constant carrying a formal uncertainty.  (v) Streams whose
activity derives from $10^{3}$--$10^{4}$-yr-old material (the Taurid
complex being the extreme case, for which our young-trail component
correctly vanishes and an analytic 7:2 resonant-swarm term following
\citealt{asher1993} is used instead) are outside the scope of direct
trail simulation at these particle counts.
Within these limits the anchored model reproduces the \emph{identity} and
\emph{timing} of the documented Leonid storms --- the 1767 trail in both
2001 and 2002, to 1 and 40~min --- and their amplitudes to
$\times0.89$--$\times1.25$; the 1932 excess and the 1866 deficit are
why we nevertheless issue Leonid forecasts at class B rather than A.
The 2034 prediction is falsifiable in detail and in a way that
discriminates between the model's remaining error sources: ZHR of order
$1.4\times10^{3}$ at $03{:}30\pm1$~h~UT on November 18 from the young
1932 trail, and a second maximum of order $1.2\times10^{3}$ at
$23{:}45$~UT ($\pm3$~h) from the nine-revolution 1733 trail.  The first
sits at the young end of the validated age range and should be read with
the 1999/2000 1932 excess in mind; the second is a section no published
forecast nominates, so it tests the epoch-matched geometry of
Sect.~\ref{sec:systematics} on its own.
\section{Conclusions}
We have constructed and calibrated a dust-trail model combining
Crifo--Rodionov ejection with unit-weight Monte-Carlo size sampling, full
N-body trail evolution, and a nodal-kernel ZHR forecast with
literature-grounded cross-section, dilution, resonance-protection, and
uncertainty terms.  In validating it against the documented Leonid
storms we found that its dominant error was the parent's back-integrated
trajectory: relative to the observation-fitted apparition solutions
distributed through the JPL ephemeris, a self-consistent
back-integration of 55P/Tempel--Tuttle misplaces the 1866 ejection site
by $0.08$--$0.15$~au and every pre-1700 site by more than 1~au.
Anchoring the ejection sites to that ephemeris, while retaining our own
propagation of the dust, removes an age-proportional bias in the
modelled nodal distances and recovers the classical attribution of the
2001 and 2002 storms to the 1767 trail, with peak times accurate to 1
and 40~min and amplitudes to $\times1.25$ and $\times0.90$.  It also
withdraws the $10^4$-level 2033 storm predicted by the unanchored model,
which we now identify as an artefact of that back-integration error.
Auditing every maximum the model generates, rather than only those the
classical analyses nominate, locates the failure that remains: the
two-revolution 1932 trail, an encounter the classical analyses also
predict and observers did detect, is over-predicted eighteenfold, while
four- to five-revolution sections are under-predicted two- to
six-fold.  Against Table~I of \citet{lyytinen2000} our mean-anomaly
factor is accurate to 3--13\% at seven revolutions and $1.7$ times too
large at two, and our old-trail nodes are displaced
$1$--$4\times10^{-3}$~au too far inside Earth's orbit.
A second systematic emerged from the forecast itself and is independent
of the parent: a dataset read at an epoch other than its own holds every
trail's node where the snapshot left it, and over 0.12 revolutions the
real nodes move a median $1.2\times10^{-2}$~au, $23\sigma_r$, under
Jovian perturbation the frozen elements cannot follow --- so each
forecast year must be computed from a dataset snapshotted in that year.
The two systematics are of a kind: both are errors of \emph{where the
dust is put}, invisible to convergence tests, and both were found only
by comparing computations that should have agreed.
For the coming Tempel--Tuttle cycle the anchored, epoch-matched model
predicts three active years.  The strongest is 2034 November 18, with
two comparable maxima twenty hours apart: ZHR $\approx1.4\times10^{3}$ at
03:30~UT from the three-revolution 1932 trail and $\approx1.2\times10^{3}$
at 23:45~UT from the nine-revolution 1733 trail (class B, 0.29--3.5).
The 1932 rate is $2.6\times10^{4}$ without the empirical density factor
and $1.4\times10^{3}$ with it, so which of the two maxima dominates that
night is a clean test of the correction, since the predicted times and
node distances do not depend on it.  That section's population index is
also unmeasurable, which puts its rate between $4\times10^{2}$ and
$1.4\times10^{3}$ and can by itself reverse the ordering.
2033 November 17 carries
ZHR $\approx1.4\times10^{3}$ at 22:56~UT from the 1899 trail --- the same
trail that produced the 1999 storm --- and 2035 November 19
ZHR $\approx650$ from the 1833 and 1800 sections.  2031 and 2032 stay at
the annual background.
\section*{Data availability}

The parent element sets used here, including the full $8\times8$ 55P
covariance matrix of Table~\ref{tab:cov}, are as retrieved from the JPL
Small-Body Database (\texttt{sbdb.api?sstr=55P\&cov=mat}) and are
reproduced in full in that table.  The epoch-matched simulation datasets
underlying Tables~\ref{tab:leo9802}--\ref{tab:cmp} and
Figs.~\ref{fig:year1999}--\ref{fig:year2035} are archived at Zenodo
\citep{abe2026data}, \url{https://doi.org/10.5281/zenodo.22004211}; each
carries its generating parameters --- grain count, ejection era, epoch,
force model, anchoring flag --- in its own header, so that any quoted
result can be traced to
the run that produced it.  The same deposit holds the orbit-covariance
ensemble of Sect.~\ref{sec:systematics} member by member, each with its
drawn orbit, its trail snapshot and its measured node, so that it can be
re-measured without repeating the integrations, and the realisation and
grain-count studies of the 1999 anchor (Sect.~\ref{sec:calib}).  Every
file is plain JSON, its schema documented in the deposit, so reading one
requires no particular software; the same files are at the same time the
native format of \textsc{meteorium}, a free viewer distributed through the
App Store (\url{https://apps.apple.com/app/id6798546441}), which opens a
downloaded dataset and its forecast bundle directly, so that a simulation
can be examined interactively rather than only read as numbers.  The
implementation itself is not distributed.  The model is specified in
Sects.~\ref{sec:stream} and~\ref{sec:forecast}, including the numerical
value of every constant it uses, the functional form of every term, and
the ceilings and cutoffs applied to each.

\section*{CRediT authorship contribution statement}
\textbf{Shinsuke Abe:} Conceptualization, Methodology, Software,
Validation, Formal analysis, Investigation, Data curation, Writing ---
original draft, Writing --- review and editing, Visualization, Project
administration.

\section*{Declaration of competing interest}
The author declares no known competing financial interests or personal
relationships that could have appeared to influence the work reported in
this paper.

\section*{Funding}
This research did not receive any specific grant from funding agencies in
the public, commercial, or not-for-profit sectors.

\section*{Acknowledgements}
The stream integrations use \textsc{rebound} and \textsc{reboundx}.
Shower parameters follow the IMO Meteor Shower Calendar and the IAU
Meteor Data Center.
\section*{Declaration of generative AI and AI-assisted technologies in
the manuscript preparation process}
During the preparation of this work the author used Claude (Anthropic) in
order to improve the language and readability of the manuscript, including
condensing the abstract to the journal's length limits.
After using this tool the author reviewed and edited the content as needed
and takes full responsibility for the content of the published article.


\begin{thebibliography}{99}
\bibitem[Abe et al.(2003)]{abe2003}
Abe, S., Yano, H., Ebizuka, N., Sugimoto, M., Kasuga, T., Watanabe,
J.-I. 2003. Twin peaks of the 2002 Leonid meteor storm observed in the
Leonid MAC airborne mission. PASJ 55, 559--565.
\bibitem[Abe(2026)]{abe2026data}
[dataset] Abe, S. 2026. Leonid dust trails in the 2030s: stream
simulations, parent element sets and encounter forecasts, version
leonids-2030s-v1. Zenodo. https://doi.org/10.5281/zenodo.22004211.
\bibitem[Agarwal et al.(2016)]{agarwal2016}
Agarwal, J., A'Hearn, M.~F., Vincent, J.-B., et al. 2016.
Acceleration of individual, decimetre-sized aggregates in the lower coma
of comet 67P/Churyumov--Gerasimenko. MNRAS 462, S78--S88.
\bibitem[Arlt(1998)]{arlt1998}
Arlt, R. 1998. Bulletin 13 of the International Leonid Watch: the 1998
Leonid meteor shower. WGN, J.\ IMO 26, 239--248.
\bibitem[Arlt and Gyssens(2000)]{arlt2000}
Arlt, R., Gyssens, M. 2000. Bulletin 16 of the International Leonid
Watch: results of the 2000 Leonid meteor shower. WGN, J.\ IMO 28,
195--208.
\bibitem[Arlt et al.(1999)]{arlt1999}
Arlt, R., Bellot Rubio, L., Brown, P., Gyssens, M. 1999.
Bulletin 15 of the International Leonid Watch: first global analysis of
the 1999 Leonid storm. WGN, J.\ IMO 27, 286--295.
\bibitem[Arlt et al.(2001)]{arlt2001}
Arlt, R., Kac, J., Krumov, V., Buchmann, A., Verbert, J. 2001.
Bulletin 17 of the International Leonid Watch: first global analysis of
the 2001 Leonid storms. WGN, J.\ IMO 29, 187--194.
\bibitem[Arlt et al.(2002)]{arlt2002}
Arlt, R., Krumov, V., Buchmann, A., Kac, J., Verbert, J. 2002.
Bulletin 18 of the International Leonid Watch: preliminary analysis of
the 2002 Leonid meteor shower. WGN, J.\ IMO 30, 205--212.
\bibitem[Asher(1999)]{asher1999}
Asher, D.~J. 1999. The Leonid meteor storms of 1833 and 1966.
MNRAS 307, 919--924.
\bibitem[Asher(2000)]{asher1999imc}
Asher, D.~J. 2000. Leonid dust trail theories. In: Proc.\ International
Meteor Conference 1999, Frasso Sabino, 5--21.
\bibitem[Asher et al.(1999)]{asher1999res}
Asher, D.~J., Bailey, M.~E., Emel'yanenko, V.~V. 1999. Resonant
meteoroids from Comet Tempel--Tuttle in 1333: the cause of the
unexpected Leonid outburst in 1998. MNRAS 304, L53--L56.
\bibitem[Asher \& Clube(1993)]{asher1993}
Asher, D.~J., Clube, S.~V.~M. 1993. An extraterrestrial influence during
the current glacial--interglacial. QJRAS 34, 481--511.
\bibitem[Brown et al.(2002)]{brown2002}
Brown, P., Campbell, M.D., Hawkes, R.L., Theijsmeijer, C., Jones, J.
2002. Multi-station electro-optical observations of the 1999 Leonid
meteor storm. Planet.\ Space Sci.\ 50, 45--55.
\bibitem[Brown \& Jones(1998)]{brown1998}
Brown, P., Jones, J. 1998. Simulation of the formation and evolution of
the Perseid meteoroid stream. Icarus 133, 36--68.
\bibitem[Burns et al.(1979)]{burns1979}
Burns, J.~A., Lamy, P.~L., Soter, S. 1979. Radiation forces on small
particles in the solar system. Icarus 40, 1--48.
\bibitem[Crifo \& Rodionov(1997)]{crifo1997}
Crifo, J.~F., Rodionov, A.~V. 1997. The dependence of the circumnuclear
coma structure on the properties of the nucleus. Icarus 127, 319--353.
\bibitem[Della Corte et al.(2015)]{dellacorte2015}
Della Corte, V., Rotundi, A., Fulle, M., et al. 2015. GIADA: shining a
light on the monitoring of the comet dust production from the nucleus of
67P/Churyumov--Gerasimenko. A\&A 583, A13.
\bibitem[Egal(2020)]{egal2020review}
Egal, A. 2020. Forecasting meteor showers: a review. Planet.\ Space
Sci.\ 185, 104895.
\bibitem[Egal et al.(2019)]{egal2019}
Egal, A., Wiegert, P., Brown, P.~G., et al. 2019. Meteor shower
modeling: past and future Draconid outbursts. Icarus 330, 123--141.
\bibitem[Egal et al.(2020)]{egal2020}
Egal, A., Brown, P.~G., Rendtel, J., Campbell-Brown, M., Wiegert, P.
2020. Modeling the past and future activity of the Halleyid meteor
showers. A\&A 642, A120.
\bibitem[Egal et al.(2023)]{egal2023}
Egal, A., Wiegert, P., Brown, P.~G., Vida, D. 2023. Modeling the 2022
tau-Herculid outburst. ApJ 949, 96.
\bibitem[G\"ockel \& Jehn(2000)]{gockel2000}
G\"ockel, C., Jehn, R. 2000. Testing cometary ejection models to fit the
1999 Leonids and to predict future showers. MNRAS 317, L1--L5.
\bibitem[Jacchia et al.(1967)]{jacchia1967}
Jacchia, L.~G., Verniani, F., Briggs, R.~E. 1967. An analysis of the
atmospheric trajectories of 413 precisely reduced photographic meteors.
Smithson.\ Contrib.\ Astrophys.\ 10, 1--139.
\bibitem[Jenniskens(2006)]{jenniskens2006}
Jenniskens, P. 2006. Meteor Showers and their Parent Comets. Cambridge
University Press, Cambridge.
\bibitem[Jenniskens et al.(2000)]{jenniskens2000}
Jenniskens, P., Crawford, C., Butow, S.~J., et al. 2000. Lorentz shaped
comet dust trail cross section from new hybrid visual and video meteor
counting technique. Earth Moon Planets 82--83, 191--208.
\bibitem[Kokhirova \& Borovi\v{c}ka(2011)]{kokhirova2011}
Kokhirova, G.~I., Borovi\v{c}ka, J. 2011. Observations of the 2009
Leonid activity by the Tajikistan fireball network. A\&A 533, A115.
\bibitem[Kondrat'eva \& Reznikov(1985)]{kondrateva1985}
Kondrat'eva, E.~D., Reznikov, E.~A. 1985. Comet Tempel--Tuttle and the
Leonid meteor swarm. Sol.\ Syst.\ Res.\ 19, 96--101.
\bibitem[Koten et al.(2011)]{koten2011}
Koten, P., Borovi\v{c}ka, J., Kokhirova, G.~I. 2011. Activity of the
Leonid meteor shower on 2009 November 17. A\&A 528, A94.
\bibitem[Langbroek(2002)]{langbroek2002}
Langbroek, M. 2002. Observational evidence for `punctuated equilibria'
in the evolution of Leonid dust trail widths and implications for
meteor rate predictions. MNRAS 334, L16--L20.
\bibitem[Lyytinen and Van Flandern(2000)]{lyytinen2000}
Lyytinen, E.J., Van Flandern, T. 2000. Predicting the strength of Leonid
outbursts. Earth Moon Planets 82--83, 149--166.
\bibitem[Lyytinen et al.(2001)]{lyytinen2001}
Lyytinen, E., Nissinen, M., Van Flandern, T. 2001. Improved 2001 Leonid
storm predictions from a refined model. WGN, J.\ IMO 29, 110--118.
\bibitem[Marsden et al.(1973)]{marsden1973}
Marsden, B.~G., Sekanina, Z., Yeomans, D.~K. 1973. Comets and
nongravitational forces. V. AJ 78, 211--225.
\bibitem[Maslov(2007)]{maslov2007}
Maslov, M. 2007. Leonid predictions for the period 2001--2100.
WGN, J.\ IMO 35, 5--12.
\bibitem[McNaught and Asher(1999b)]{mcnaught1999b}
McNaught, R.H., Asher, D.J. 1999b. Variation of Leonid maximum times with
location of observer. Meteorit.\ Planet.\ Sci.\ 34, 975--978.
\bibitem[McNaught \& Asher(1999)]{mcnaught1999}
McNaught, R.~H., Asher, D.~J. 1999. Leonid dust trails and meteor
storms. WGN, J.\ IMO 27, 85--102.
\bibitem[McNaught \& Asher(2002)]{mcnaught2002}
McNaught, R.~H., Asher, D.~J. 2002. Leonid dust trail structure and
predictions for 2002. WGN, J.\ IMO 30, 132--143.
\bibitem[Rein \& Liu(2012)]{rein2012}
Rein, H., Liu, S.-F. 2012. REBOUND: an open-source multi-purpose N-body
code for collisional dynamics. A\&A 537, A128.
\bibitem[Rein \& Spiegel(2015)]{rein2015}
Rein, H., Spiegel, D.~S. 2015. IAS15: a fast, adaptive, high-order
integrator for gravitational dynamics. MNRAS 446, 1424--1437.
\bibitem[Rendtel et al.(2000)]{rendtel2000}
Rendtel, J., Betlem, H., Ter Kuile, C., Lyytinen, E., Jenniskens, P.,
Brown, P. 2000. Video observations of the 1999 Leonid storm from Jordan.
WGN, J.\ IMO 28, 20--24.
\bibitem[Sato \& Watanabe(2007)]{sato2007}
Sato, M., Watanabe, J. 2007. Origin of the 2006 Orionid outburst.
PASJ 59, L21--L24.
\bibitem[Tamayo et al.(2020)]{tamayo2020}
Tamayo, D., Rein, H., Shi, P., Hernandez, D.~M. 2020. REBOUNDx: a
library for adding conservative and dissipative forces to otherwise
symplectic N-body integrations. MNRAS 491, 2885--2901.
\bibitem[Vaubaillon et al.(2005a)]{vaubaillon2005}
Vaubaillon, J., Colas, F., Jorda, L. 2005a. A new method to predict
meteor showers. I. Description of the model. A\&A 439, 751--760.
\bibitem[Vaubaillon et al.(2005b)]{vaubaillon2005b}
Vaubaillon, J., Colas, F., Jorda, L. 2005b. A new method to predict
meteor showers. II. Application to the Leonids. A\&A 439, 761--770.
\bibitem[Vida et al.(2022)]{vida2022}
Vida, D., Blaauw Erskine, R.~C., Brown, P.~G., et al. 2022. Computing
optical meteor flux using Global Meteor Network data. MNRAS 515,
2322--2339.
\bibitem[Vida et al.(2024)]{vida2024}
Vida, D., Segon, D., Sato, M., et al. 2024. A global observational
analysis of the 2023 lambda-Sculptorid meteor shower outburst.
arXiv:2402.07769.
\bibitem[Watanabe \& Sato(2008)]{watanabe2008}
Watanabe, J., Sato, M. 2008. Activities of parent comets and related
meteor showers. Earth Moon Planets 102, 111--116.
\bibitem[Whipple(1951)]{whipple1951}
Whipple, F.~L. 1951. A comet model. II. Physical relations for comets
and meteors. ApJ 113, 464--474.
\end{thebibliography}
\end{document}